\pdfoutput=1
\documentclass[conference]{IEEEtran}
\ifCLASSINFOpdf
  \usepackage[pdftex]{graphicx}
\else
  \usepackage[dvips]{graphicx}
\fi
\usepackage[cmex10]{amsmath}
\usepackage{amssymb}

\usepackage{url}

\newtheorem{theorem}{Theorem}
\newtheorem{lemma}[theorem]{Lemma}
\newtheorem{fact}[theorem]{Fact} 
\newtheorem{example}{Example}

\newcommand{\todo}[1]{{$\langle\!\langle$\marginpar[$\blacksquare$]{$\blacksquare$}\textsf{#1}$\rangle\!\rangle$}}

\newcommand{\set}[1]{\left\{ #1 \right\}}
\newcommand{\Set}[1]{\set{#1}}

\newcommand{\Setdef}[2]{\left\{{#1}\mid{#2}\right\}}
\newcommand{\Tuple}[1]{\ensuremath{\left\langle{#1}\right\rangle}}

\newcommand{\Card}[1]{\ensuremath{{\left|{#1}\right|}}}
\newcommand{\BigOh}{\mathcal{O}}
\newcommand{\Tool}[1]{\textsf{\small #1}}
\newcommand{\STool}[1]{\textsf{\scriptsize #1}}

\newcommand{\Booleans}{\mathbb{B}}

\newcommand{\T}{\top}
\newcommand{\F}{\bot}

\newcommand{\X}{\oplus}
\newcommand{\XX}{\,{\oplus}\,}
\newcommand{\Def}{\mathrel{:=}}

\newcommand{\VarsOf}[1]{\operatorname{vars}(#1)}
\newcommand{\LitsOf}[1]{\operatorname{lits}(#1)}

\newcommand{\orpart}{\phi_\textup{or}}
\newcommand{\xorpart}{\phi_\textup{xor}}

\newcommand{\XC}{D}

\newcommand{\XCI}[1]{D_{#1}}
\newcommand{\XCB}{E}

\begin{document}
%
\title{Extending Clause Learning SAT Solvers with Complete Parity Reasoning (extended version)}

\author{\IEEEauthorblockN{Tero Laitinen, Tommi Junttila, and Ilkka Niemel\"a}
\IEEEauthorblockA{Aalto University\\
  Department of Information and Computer Science\\
  PO Box 15400, FI-00076 Aalto, Finland\\
  Email: \{Tero.Laitinen,Tommi.Junttila,Ilkka.Niemela\}@aalto.fi}
}


%


\maketitle

\begin{abstract}
Instances of logical cryptanalysis, circuit verification, and bounded
model checking can often be succinctly represented as a combined
satisfiability (SAT) problem where an instance is a combination of
traditional clauses and parity constraints. 
This paper studies how such combined problems can be efficiently solved by
augmenting a modern SAT solver with an xor-reasoning module in the DPLL(XOR)
framework.
A new xor-reasoning module that deduces all possible implied literals
using incremental Gauss-Jordan elimination is presented.
A decomposition technique that can greatly reduce the size of parity
constraint matrices while allowing still to deduce all implied literals is
presented.
It is shown how to eliminate variables occuring only in parity
constraints while preserving the decomposition. The proposed techniques are
evaluated experimentally.
\end{abstract}


%
\IEEEpeerreviewmaketitle

\section{Introduction}

Propositional satisfiability (SAT) solvers (see e.g.~\cite{Handbook:CDCL})
provide a powerful solution technique in many industrial application domains.
Representing an instance of propositional satisfiability in conjunctive
normal form (CNF) allows very efficient Boolean constraint propagation
and conflict-driven clause learning (CDCL) techniques. 
However, CNF-based solvers can scale poorly on instances consisting on
straightforward CNF-encoding of parity (xor) constraints
\cite{Urquhart:JACM1987}.
Such xor-constraints occur frequently in domains such as logical cryptanalysis,
circuit verification, and bounded model checking.
Considering this and
recalling that an instance consisting only of xor-constraints can be
solved in polynomial time using Gaussian elimination,
it is no wonder that many approaches for combining CNF-level and
xor-constraint reasoning have been presented~\cite{Li:AAAI2000,Li:IPL2000,BaumgartnerMassacci:CL2000,Li:DAM2003,HeuleMaaren:SAT2004,HeuleEtAl:SAT2004,Chen:SAT2009,SoosEtAl:SAT2009,LJN:ECAI2010,Soos,LJN:ICTAI2011,LJN:SAT2012}.
These approaches extend CNF-level SAT solvers by
implementing different forms of constraint propagation for xor-constraints,
ranging from plain unit propagation
via equivalence reasoning
to Gaussian elimination.
Compared to unit propagation,
which has efficient implementation techniques,
equivalence reasoning and Gaussian elimination allow stronger propagation but
are computationally much more costly.
%

In this paper we make two contributions in this field.
First,
we present an xor-reasoning technique based on Gauss-Jordan elimination that
provides complete constraint propagation for xor-constraints
in the following sense:
Given a conjunction $\xorpart$ of xor-constraints and
values for some of its variables (so-called xor-assumptions provided by the CNF-level master search engine),
the module can
(i) \emph{decide} whether $\xorpart$ is satisfiable under the xor-assumptions,
and
(ii) find \emph{all} the literals and equivalences implied by $\xorpart$ and the xor-assumptions.
This is better than
(i) equivalence reasoning which cannot always decide the satisfiability or find all the implied literals,
and
(ii) Gaussian elimination which can decide satisfiability but not necessarily finds all the implied literals (as illustrated in Sect.~\ref{Sect:GaussJordan}).%
\footnote{We've just learned that the use of Gauss-Jordan has also been independently discovered in \cite{HanJiang:CAV2012}: the main difference to our work is that we (i) do not consider Craig interpolants but (ii) can also find all the implied equivalences.}

Our second contribution is a new decomposition theorem that sometimes allows
us to split the xor-constraint part $\xorpart$ into components that can
be handled individually.
This technique supersedes the well-known ``connected components'' approach
that exploits variable disjoint components of $\xorpart$.
Instead, we use a variant of ``biconnected components'' by splitting
$\xorpart$ into components that can be connected to each other only by
single cut variables.
We prove that if we can provide full propagation for each of the components,
we have full propagation for the whole xor-part $\xorpart$ as well.
We show how the structure of biconnected components can be preserved while
eliminating most of the variables occurring only in the xor-part leading to
more compact representation of the formula.
The presented xor-reasoning, decomposition, and variable elimination techniques
are evaluated experimentally on large sets of benchmark instances.

The proofs of Lemmas and Theorems can be found in the appendix.
\section{Preliminaries}

\newcommand{\Equal}{\equiv}
\newcommand{\Equiv}{\equiv}
\newcommand{\Models}{\models}

\newcommand{\Var}{x}
\newcommand{\AL}{\tilde{l}}
\newcommand{\IL}{\hat{l}}
\newcommand{\parity}[1]{p_{#1}}
\newcommand{\TA}{\tau}

Let $\Booleans = \Set{\F,\T}$ be the set of truth values ``false'' and ``true''.
A literal is a Boolean variable $x$ or its negation $\neg x$
(as usual, $\neg \neg x$ will mean $x$),
and a clause is a disjunction of literals.
If $\phi$ is any kind of formula or equation,
(i) $\VarsOf{\phi}$ is the set of variables occurring in it,
(ii) $\LitsOf{\phi} = \Setdef{x,\neg x}{x\in\VarsOf{\phi}}$ is the set of literals over $\VarsOf{\phi}$,
and
(iii) a truth assignment for $\phi$ is a, possibly partial,
function $\TA : \VarsOf{\phi} \to \Booleans$.
A truth assignment satisfies (i) a variable $x$ if $\TA(x)=\T$,
(ii) a literal $\neg x$ if $\TA(x)=\F$, and
(iii) a clause $(l_1 \lor .. \lor l_k)$ if it satisfies at least one literal $l_i$ in the clause.

An \emph{xor-constraint} is an equation of form $x_1 \X ... \X x_k \Equal p$,
where the $x_i$s are 
Boolean variables and
$p \in \Booleans$ is the parity.%
\footnote{The correspondence of xor-constraints to the ``xor-clause'' representation used e.g.~in \cite{LJN:ECAI2010,LJN:ICTAI2011,LJN:SAT2012} is straightforward: $x_1 \X ... \X x_k \Equal \T$ corresponds to the xor-clause $(x_1 \X ... \X x_k)$ and $x_1 \X ... \X x_k \Equal \F$ to $(x_1 \X ... \X x_k \X \T)$.}
We implicitly assume that duplicate variables are always removed from
the equations,
e.g.~$x_1 \X x_2 \X x_1 \X x_3 \Equal \T$ is always simplified into
$x_2 \X x_3 \Equal \T$.
If the left hand side does not have variables, 
then it equals to $\F$;
the equation $\F \Equal \T$ is a contradiction and
$\F \Equal \F$ a tautology.
We identify the xor-constraint $\Var \Equal \T$ with the literal $\Var$ and
$\Var \Equal \F$ with $\neg\Var$.
A truth assignment $\TA$ satisfies an xor-constraint $(x_1 \X ... \X x_k \Equal p)$ if $\TA(x_1) \X ... \X \TA(x_k) = p$.
%
%
%

A \emph{cnf-xor formula} is a conjunction $\orpart \land \xorpart$,
where
$\orpart$ is a conjunction of clauses
and
$\xorpart$ is a conjunction of xor-constraints.
A truth assignment satisfies $\orpart \land \xorpart$ if it satisfies
every clause and xor-constraint in it.

%
%
\subsection{DPLL(XOR) and Xor-Reasoning Modules}

We are interested in solving the satisfiability of cnf-xor formulas of
the form $\orpart \land \xorpart$ defined above.
%
%
Similarly to the DPLL($T$) approach for Satisfiability Modulo
Theories, see e.g.~\cite{NieuwenhuisEtAl:JACM06,Handbook:SMT},
the DPLL(XOR) approach \cite{LJN:ECAI2010} for solving cnf-xor formulas
consists of
(i) a conflict-driven clause learning (CDCL) SAT solver that takes care of solving the CNF-part $\orpart$,
and
(ii) an \emph{xor-reasoning module} that handles the xor-part $\xorpart$.
The CDCL solver is the master process,
responsible of guessing values for the variables according to some heuristics
(``branching''),
performing propagation in the CNF-part, conflict analysis, restarts etc.
The xor-reasoning module receives variable values,
called xor-assumptions,
from the CDCL solver and
checks
(i) whether the xor-part can still be satisfied under the xor-assumptions,
and
(ii) whether some variable values, called xor-implied literals,
are implied by the xor-part and the xor-assumptions.
These checks can be incomplete,
like in~\cite{LJN:ECAI2010,LJN:ICTAI2011} for the satisfiability and
in~\cite{LJN:ECAI2010,LJN:ICTAI2011,SoosEtAl:SAT2009} for the implication checks,
as long as the satisfiability check is complete when all the variables have values.
%

The very basic interface for an xor-reasoning module can consist of the following methods:
\begin{itemize}
\item
  $\operatorname{init}(\xorpart)$
  initializes the module with $\xorpart$.
  It may return ``unsat'' if it finds $\xorpart$ unsatisfiable,
  or a set of \emph{xor-implied literals},
  i.e.~literals $\IL$ such that $\xorpart \Models \IL$ holds.

\item
  $\operatorname{assume}(\AL)$ is used to communicate a new variable value
  $\AL$ deduced in the CNF solver part to the xor-reasoning module.
  This value, called \emph{xor-assumption} literal $\AL$,
  is added to the list of current xor-assumptions.
  If $[\AL_1,...,\AL_k]$ are the current xor-assumptions,
  the module then tries to 
  (i) deduce whether $\xorpart \land \AL_1 \land ... \land \AL_k$ became unsatisfiable, i.e.~whether an \emph{xor-conflict} was encountered,
  and if this was not the case,
  (ii) find \emph{xor-implied literals},
  i.e.~literals $\IL$ for which $\xorpart \land \AL_1 \land ... \land \AL_k \Models \IL$ holds.
  The xor-conflict or the xor-implied literals are then returned to the CNF solver part so that it can start conflict analysis (in the case of xor-conflict) or
  extend its current partial truth assignment with the xor-implied literals.

  In order to facilitate conflict-driven backjumping and clause learning
  in the CNF solver part,
  the xor-reasoning module has to provide a clausal \emph{explanation} for each
  xor-conflict and xor-implied literal it reports.
  That is,
  \begin{itemize}
  \item if $\xorpart \land \AL_1 \land ... \land \AL_k$ is
  deduced to be unsatisfiable,
  then the module must report a (possibly empty) clause
  $({\neg l'_1} \lor ... \lor {\neg l'_m})$
  such that
  (i) each $l'_i$ is an xor-assumption or an xor-implied literal,
  and
  (ii) $\xorpart \land l'_1 \land ... \land l'_m$ is unsatisfiable
  (i.e.~$\xorpart \Models ({\neg l'_1} \lor ... \lor {\neg l'_m})$);
  and
  \item
  if it was deduced that
  $\xorpart \land \AL_1 \land ... \land \AL_k \Models \IL$ for some $\IL$,
  then the module must report a clause 
  $({\neg l'_1} \lor ... \lor {\neg l'_m} \lor \IL)$ such that
  (i) each $l'_i$ is an xor-assumption or an xor-implied literal reported earlier,
  and
  (ii)
  $\xorpart \land l'_1 \land ... \land l'_m \Models \IL$,
  i.e.~$\xorpart \Models ({\neg l'_1} \lor ... \lor {\neg l'_m} \lor \IL)$. 
  \end{itemize}
\item
  $\operatorname{backtrack}()$ retracts the latest xor-assumption
  and
  all the xor-implied literals deduced after it.
\end{itemize}
Naturally, variants of this interface are easily conceivable.
For instance, a larger set of xor-assumptions can be given with
the $\operatorname{assume}$ method at once instead of only one.

For xor-reasoning modules based on equivalence reasoning,
see~\cite{LJN:ECAI2010,LJN:ICTAI2011}.
The Gaussian elimination process in~\cite{SoosEtAl:SAT2009,Soos}
can also be easily seen as an xor-reasoning module.

\section{Incremental Gauss-Jordan Elimination}
\label{Sect:GaussJordan}

We now develop an xor-reasoning technique that can,
given a conjunction $\xorpart$ of xor-constraints and
a conjunction $\AL_1 \land ... \land \AL_k$ of xor-assumption literals,
(i) \emph{decide} whether $\xorpart \land \AL_1 \land ... \land \AL_k$ is satisfiable
or not,
and
(ii) if it is, to find \emph{all} the literals and equivalences implied by
$\xorpart \land \AL_1 \land ... \land \AL_k$.
The proposed technique can be seen as an incremental, Boolean-level version of
the Gauss-Jordan elimination process,
or a Boolean-level variant of the linear arithmetic solver
described in~\cite{DutertreMoura:CAV2006}.

Before going into the details,
let us first briefly note why Gaussian elimination,
used e.g.\ in Cryptominisat~\cite{SoosEtAl:SAT2009,Soos} version 2.9.2,
is not enough to find all the implied literals
(although it can detect unsatisfiability perfectly).
Basically, the reason is that Gaussian elimination presents the xor-constraints in $\xorpart$
with a row echelon form matrix, where pivoting upwards is not performed.
As an example,
consider the row echelon form matrix-like representation
\[
 \begin{array}{r@{}r@{}r@{}r@{}r@{}r@{}r}
  x_1 & \X x_2 &        & \X x_4 &        & {}\Equal{} & \T\\
      &    x_2 & \X x_3 &        & \X x_5 & {}\Equal{} & \F\\
      &       &     x_3 & \X x_4 & \X x_5 & {}\Equal{} & \T
 \end{array}
\]
for a conjunction $\xorpart$ of xor-constraints.
It is easy to deduce from this that $\xorpart$ is satisfiable
but
not that $x_1$ must always be false, i.e.~that $\xorpart \Models {x_1 \Equiv \F}$.
%

\subsection{Tableaux i.e.~Reduced Row Echelon Form Matrices}

\newcommand{\Eqs}{\mathcal{E}}

We begin by giving an equation form representation and
the basic operations we need for reduced row echelon matrices.
A \emph{tableau} for a satisfiable conjunction $\xorpart$ of xor-constraints
is a set $\Eqs$ of equations of form
$\Var_i \Def {\Var_{i,1} \X ... \X \Var_{i,k_i} \X \parity{i}}$,
where
$\Var_i$,$\Var_{i,1}$,..., $\Var_{i,k_i}$ are distinct variables
in $\xorpart$
and
$\parity{i} \in \Booleans$. 
%
%
Furthermore,
it is required that
\begin{enumerate}
\item
  each variable $\Var \in \VarsOf{\xorpart}$ occurs \emph{at most once}
  as the left hand side variable in the equations in $\Eqs$,
\item
  if a variable $\Var \in \VarsOf{\xorpart}$ occurs as
  the left hand side variable in an equation,
  then it does not occur in the right hand side of any equation,
  and
\item
  $\bigwedge_{{\Var_i \Def \Var_{i,1} \X ... \X \Var_{i,k_i} \X \parity{i}} \in \Eqs}
   (\Var_i \X \Var_{i,1} \X ... \X \Var_{i,k_i} \Equal \parity{i})$
  is logically equivalent to $\xorpart$.
\end{enumerate}
The variables of $\xorpart$
occurring as left hand side variables in the equations are
called \emph{basic variables} while
the others are \emph{non-basic variables} in $\Eqs$.
If $\Eqs$ has $n$ non-basic variables,
then $\xorpart$ has $2^n$ satisfying truth assignments.
Observe that
a tableau can be seen as a linear arithmetic modulo 2 matrix equation;
under a variable order where basic variables are first,
the matrix will be in the reduced row echelon form.
\begin{example}
  Take the conjunction
  $(a \X c \X e \Equal \T) \land (a \X b \X d \X e \Equal \T)$.
  A tableau for it is
  $\left\{\begin{array}{rcl@{}l@{}l@{}l@{}l}
  a & \Def & c &      &  & \X e & \X\T\\
  b & \Def & c & \X d &  &      & \X\F\\
  \end{array}\right\}$,
  or
  $\left(\begin{smallmatrix}
    1 & 0 & 1 & 0 & 1\\
    0 & 1 & 1 & 1 & 0\\
  \end{smallmatrix}\right)
  \left(\begin{smallmatrix}a & b & c & d & e\end{smallmatrix}\right)^\textup{T} = 
    \left(\begin{smallmatrix} 1 \\0
    \end{smallmatrix}\right)$
  as a matrix equation;
  the first matrix is in the reduced row echelon form.
\end{example}

Given a conjunction $\xorpart = \XCI1 \land ... \land \XCI{m}$ of xor-constraints,
it is easy to build a tableau for it (or to detect that the conjunction
is unsatisfiable, in which case it does not have a tableau).
We start with the empty tableau,
and for each xor-constraint $\XC$ in the conjunction apply the following:
\begin{enumerate}
\item
  Eliminate each basic variable $\Var_i$ in $\XC$ by substituting it with
  the right hand side of the equation
  $\Var_i \Def {\Var_{i,1} \X ... \X \Var_{i,k_i} \X \parity{i}}$
  already in the tableau,
  then simplify the resulting xor-constraint.
\item
  (i)
  If the resulting xor-constraint is $(\F \Equal \F)$,
  then 
  $\XC$ is a linear combination of
  the xor-constraints already 
  in the tableau and nothing is added in the tableau.

  (ii)
  If the resulting xor-constraint is 
  $(\F \Equal \T)$,
  then 
  $\XC$ is contradicting the xor-constraints already
  in the tableau and the conjunction $\xorpart$ is unsatisfiable.

  (iii)
  Otherwise,
  all the variables in the resulting xor-constraint  $(y_1 \X y_2 \X ... \X y_k \Equal \parity{})$
  are non-basic variables.
  Pick one of these variables, say $y_1$,
  insert the equation $y_1 \Def {y_2 \X ... \X y_k \X \parity{}}$
  in the tableau,
  eliminate $y_1$ from the right hand sides of other equations
  by substituting it with $y_2 \X ... \X y_k \X \parity{}$,
  and simplify the right hand sides of the equations.
\end{enumerate}
\begin{example}\label{Ex:Tableau}
  Take again the satisfiable conjunction
  $(a \X c \X e \Equal \T) \land (a \X b \X d \X e \Equal \T)$.
  When inserting $(a \X c \X e \Equal \T)$ into the empty tableau,
  we may select $a$ to be the basic variable and
  get the tableau
  $\left\{a \Def c \X e\X \T\right\}$.
  Next inserting $(a \X b \X d \X e \Equal \T)$,
  we first substitute $a$ with its definition $c \X e \X \T$,
  get $(b \X c \X d \Equal \F)$,
  select $b$ to be a basic variable,
  and obtain the tableau
  $\left\{a \Def {c \X e \X \T}, b \Def {c \X d \X \F} \right\}$.
\end{example}
%

\newcommand{\Eq}{e}
\newcommand{\Swap}[3]{\operatorname{swap}(#1,#2,#3)}

In the following,
we must be able to transform a basic variable into a non-basic one.
To do this, we must make a non-basic variable basic.
%
%
If $\Var$ is a basic variable with the equation
$\Var \Def {y_1 \X ... \X y_i \X ... \X y_k \X \parity{}}$
in a tableau $\Eqs$,
we define $\Swap{\Eqs}{x}{y_i}$ to be the tableau 
obtained 
as follows:
\begin{enumerate}
\item
  remove $\Var \Def {y_1 \X ... \X y_i \X ... \X y_k \X \parity{}}$ from $\Eqs$,
\item
  add $y_i \Def {y_1 \X ... \X y_{i-1} \X \Var \X y_{i+1} \X ... \X y_k \X \parity{}}$ in $\Eqs$,
  and
\item
  remove $y_i$ from the right hand sides of the other equations by
  substituting its occurrences with
  $y_1 \X ... \X y_{i-1} \X \Var \X y_{i+1} \X ... \X y_k \X \parity{}$.
\end{enumerate}
\begin{example}
  If
  $\Eqs =
  \left\{\begin{array}{@{}rcl@{}l@{}l@{}l@{}}
      a & \Def & c &      & \X e & \X\T \\
      b & \Def & c & \X d &      & \X\F
    \end{array}\right\}$,
  then we have
  $\Swap{\Eqs}{b}{c} =
  \left\{\begin{array}{@{}rcl@{}l@{}l@{}l@{}}
      a & \Def & b & \X d & \X e & \X\T \\
      c & \Def & b & \X d &      & \X\F   
    \end{array}\right\}$.
\end{example}

%
%
\subsection{Handling Xor-Assumptions: Assigned Tableaux}

\newcommand{\ET}[2]{\Tuple{#1,#2}}

We now show how to handle xor-assumptions,
i.e.~to \emph{decide} whether $\xorpart \land \AL_1 \land ... \land \AL_k$
is still satisfiable,
and
if yes, to find \emph{all} the literals and equivalences
implied by $\xorpart \land \AL_1 \land ... \land \AL_k$.
To do these, we introduce a concept of assigned tableaux.
To facilitate easy backtracking,
i.e.~removal of xor-assumptions,
the key idea here,
similarly to \cite{DutertreMoura:CAV2006},
is to not remove variables from the tableau when new
xor-assumptions are made but handle them separately.
In this way backtracking simply amounts to retracting xor-assumptions.

Formally,
an \emph{assigned tableau} for $\xorpart$ is a pair $\Tuple{\Eqs,\TA}$
such that
(i) $\Eqs$ is a tableau for $\xorpart$, and
(ii) $\TA$ is a, usually partial, truth assignment for $\xorpart$ in which we collect the xor-assumptions and xor-implied literals.
With respect to $\Tuple{\Eqs,\TA}$,
an equation $\Var_i \Def {\Var_{i,1} \X ... \X \Var_{i,k_i} \X \parity{i}} \in \Eqs$
is \emph{propagation saturated} if it holds that
$\TA(\Var_i)$ is defined
if and only if
$\TA(\Var_{i,j})$ is defined for all $\Var_{i,j} \in \Set{\Var_i,\Var_{i,1},...,\Var_{i,k_i}}$;
the assigned tableau $\Tuple{\Eqs,\TA}$ is \emph{propagation saturated} if
each equation in it is.
An equation $\Var_i \Def {\Var_{i,1} \X ... \X \Var_{i,k_i} \X \parity{i}}$ is
\emph{inconsistent}
if
$\TA(x)$ is defined for all $x \in \Set{\Var_i,\Var_{i,1},...,\Var_{i,k_i}}$
and
$\TA(\Var_i) \neq {\TA(\Var_{i,1}) \X ... \X \TA(\Var_{i,k_i}) \X \parity{i}}$;
if the equation is not inconsistent, it is \emph{consistent}.
An assigned tableau is inconsistent if it has an inconsistent equation;
otherwise it is consistent.
A key property of a propagation saturated assigned tableau $\Tuple{\Eqs,\TA}$
is that
its consistency is in one-to-one correspondence with
the satisfiability of
$\xorpart$ under the truth assignment $\TA$:
\begin{lemma}\label{Lemma:SimplexConsistent}
  Let $\Tuple{\Eqs,\TA}$ be a propagation saturated assigned tableau for
  $\xorpart$.
  %
  %
  The formula $\xorpart \land \bigwedge_{(x \mapsto v_x) \in \TA}(x \Equal v_x)$
  is satisfiable
  if and only if
  $\Tuple{\Eqs,\TA}$ is consistent.
\end{lemma}
From a consistent, propagation saturated assigned tableau
it is also easy 
to enumerate \emph{all} the literals that are implied by the xor-constraints
and the truth assignment in the tableau:
%
%
%
\begin{lemma}\label{Lemma:SimplexImpliedUna}
  Let $\Tuple{\Eqs,\TA}$ be a consistent,
  propagation saturated assigned tableau for $\xorpart$.
  For each literal $y \Equal v_y$ it holds that
  ${\xorpart \land  \bigwedge_{(x \mapsto v_x) \in \TA}(x \Equal v_x)} \Models (y \Equal v_y)$
  if and only if
  $\TA(y) = v_y$.
\end{lemma}

\newcommand{\Restr}[2]{#1|_{#2}}

In addition to implied literals,
we can also enumerate all implied binary xor-constraints
(i.e., equalities and disequalities between variables)
as the following Lemma shows.
\begin{lemma}\label{Lemma:SimplexImpliedBin}
  Let $\Tuple{\Eqs,\TA}$ be a consistent,
  propagation saturated assigned tableau for $\xorpart$.
  For any two distinct variables $y,z$ and any $\parity{} \in \Booleans$,
  it holds that
  ${\xorpart \land  \bigwedge_{(x \mapsto v_x) \in \TA}(x \Equal v_x)} \Models (y \X z \Equal \parity{})$
  if and only if
  \begin{enumerate}
  \item
    $\TA(y)$ and $\TA(z)$ are both defined and
    ${\TA(y) \X \TA(z) = \parity{}}$,
  \item
    $\TA(y)$ and $\TA(z)$ are undefined and
    $\Eqs$ has an equation $\Eq$ of form $y \Def {... \X z \X ...}$
    such that $\Restr{\Eq}{\TA}$ is ${y \Def z \X \parity{}}$,
    where $\Restr{\Eq}{\TA}$ is the equation obtained from $\Eq$ by substituting
    the variables in it assigned by $\TA$ with their values,
  \item
    $\TA(y)$ and $\TA(z)$ are undefined and
    $\Eqs$ has an equation $\Eq$ of form $z \Def {... \X y \X ...}$
    such that $\Restr{\Eq}{\TA}$ is ${z \Def y \X \parity{}}$,
    or
  \item
    $\TA(y)$ and $\TA(z)$ are undefined and
    $\Eqs$ has two equations, $\Eq_y$ and $\Eq_z$,
    of forms $y \Def ...$ and $z \Def ...$
    such that
    $\Restr{\Eq_y}{\TA}$ is $y \Def f$,
    $\Restr{\Eq_z}{\TA}$ is $z \Def g$,
    and
    $f \X g$ equals $\parity{}$.
  \end{enumerate}
\end{lemma}
\begin{example}
  Consider the assigned tableau $\Tuple{\Eqs,\TA}$ for a $\xorpart$
  with
  $\Eqs = \{x_1 \Def {x_3 \X x_4 \X \T}, x_2 \Def {x_3 \X x_4 \X x_5 \X \T}\}$
  and
  $\TA = \{x_5 \mapsto \T\}$.
  Now $\xorpart \land  (x_5 \Equal \T) \Models (x_1 \X x_2 \Equal \T)$
  as
  $\Restr{(x_1 \Def x_3 \X x_4 \X \T)}{\TA}$ is $x_1 \Def x_3 \X x_4 \X \T$,
  $\Restr{(x_2 \Def x_3 \X x_4 \X x_5 \X \T)}{\TA}$ is
  $x_2 \Def x_3 \X x_4 \X \F$,
  and
  $(x_3 \X x_4 \X \T) \X (x_3 \X x_4 \X \F)$ equals $\T$.
\end{example}
Such implied binary xor-constraints can be used to preprocess the cnf-xor
formula and possibly also during the search;
this topic is left for future research.

%
%
\subsubsection{Making the initial assigned tableau}
If we have a tableau for $\xorpart$ (implying that $\xorpart$ is satisfiable),
we get a corresponding consistent, propagation saturated assigned tableau
$\ET{\Eqs}{\TA}$
by simply setting $\TA(\Var_i) = \parity{i}$
for each equation 
$\Var_i \Def \parity{i}$ in $\Eqs$.
%
%
\begin{example}
  For $\xorpart = (x \X y \X z \Equal \T) \land (y \X z \Equal \F)$
  we may get the tableau $\Set{x \Def \T, y \Def z \X \F}$.
  The corresponding consistent, propagation saturated assigned tableau
  is thus $\ET{\Set{x \Def \T, y \Def {z \X \F}}}{\Set{x \mapsto \T}}$.
\end{example}

\newcommand{\Extend}{\operatorname{extend}}

%
%
\subsubsection{Extending with new xor-assumptions}
We now describe the central operation of extending a
consistent, propagation saturated assigned tableau with a new xor-assumption.
%
%
Given such an assigned tableau $\Tuple{\Eqs,\TA}$
and
an xor-assumption literal ${x \Equal v}$, 
define $\Extend(\Tuple{\Eqs,\TA}, {x \Equal v})$ to be a result of the following
non-deterministic method $\operatorname{assume}(x \Equal v)$:
\begin{enumerate}
\item
  If $\TA(x) = v$, return ``sat, no new xor-implied literals''.
\item
  If $\TA(x) \neq v$,
  return $\textup{``unsat''}$.
\item
  If $x$ is a basic variable in $\Eqs$,
  update $\Eqs$ to $\Swap{\Eqs}{x}{y}$,
  where $y$ is any $\TA$-unassigned non-basic variable in the equation for $x$;
  $x$ is now a non-basic variable.
\item
  Assign $\TA(x)  = v$.
\item
  For each equation $z \Def {x \X x'_1 \X ... \X x'_m \X \parity{}}$ in $\Eqs$,
  check whether $\TA(x'_i)$ is defined for each
  variable $x'_i$ occurring in the right hand side;
  if this is the case,
  evaluate the value $v_z$ of $z$ according to the equation and
  assign $\TA(z) = v_z$.
  The literal $z \Equal v_z$ is a new xor-implied literal.
\item
  Return ``sat'' and all the new xor-implied literals found.
\end{enumerate}

\begin{example}
  Consider the consistent, propagation saturated assigned tableau
  $\Tuple{\Eqs_0,\emptyset}$,
  where
  $\Eqs_0 =
   \left\{\begin{array}{@{\,}r@{}c@{}l@{}l@{}l@{}l@{\,}}
    a & {}\Def{} & d &      & \X f & \X \T \\
    b & {}\Def{} & d & \X e &      & \X \F \\
    c & {}\Def{} & d &      & \X f & \X \F
   \end{array}\right\}$.
  To compute $\Extend(\Tuple{\Eqs_0,\emptyset}, a\Equal\T)$,
  we first make the variable $a$ non-basic by transforming $\Eqs_0$ to
  $\Eqs_1 =
   \Swap{\Eqs_0}{a}{d} = 
    \left\{\begin{array}{@{\,}r@{}c@{}l@{}l@{}l@{}l@{\,}}
    d & {}\Def{} & a &      & \X f & \X\T \\
    b & {}\Def{} & a & \X e & \X f & \X\T \\
    c & {}\Def{} & a &      &      & \X\T
    \end{array}\right\}$
  and then assign $a$ to $\T$;
  the resulting consistent, but not propagation saturated,
  assigned tableau is $\Tuple{\Eqs_1,\Set{a \mapsto \T}}$.
  To make it propagation saturated,
  we note that $c \Def a \X \T$ has all its right hand side variables assigned
  and
  deduce a value for $c$,
  resulting in $\Tuple{\Eqs_1,\Set{a \mapsto \T, c \mapsto \F}}$.
\end{example}

%
%
\subsubsection{Backtracking}
Now observe the following: once an equation has all its variables assigned,
it will not be modified in the subsequent calls of the $\operatorname{assume}$
method until some of the variable values are retracted with the $\operatorname{backtrack}$ method.
And when this happens, at least two variables lose their values so the equation stays propagation saturated.
As a consequence, the tableau does not have to be modified when backtracking.

%
%
\subsubsection{Clausal Explanations}
Let us study how the clausal explanations for xor-conflicts
(step 2 in $\operatorname{assume}$)
and
xor-implied literals (step 5) are obtained.
\begin{itemize}
\item
  Under the reasonable assumption that the CNF solver does not make contradictory truth assignments, an xor-conflict can only happen when the xor-assumption $x \Equal v$ is an xor-implied literal derived earlier but ignored so far for some scheduling reason by the CNF-part solver.
  Thus there is an equation $x \Def y_1 \X ... \X y_m \X \parity{}$ in $\Eqs$
  such that
  $\TA(x) \Def \TA(y_1) \X ... \X \TA(y_m) \X \parity{}$ and $\TA(x) \neq v$;
  the explanation is now the clause
  ${\neg(y_1 \Equal \TA(y_1))} \lor ... \lor \neg(y_m \Equal \TA(y_m)) \lor \neg(x \Equal v)$.
\item
  For an xor-implied literal $z \Equal v_z$ derived in step 5,
  the explanation is simply a clause in the straightforward CNF translation
  of the equation,
  i.e.~$\neg(x\Equal\TA(x)) \lor \neg({x'_1\Equal\TA(x'_1)}) \lor ... \lor \neg(x'_m\Equal\TA(x'_m)) \lor (z \Equal v_z)$.
\end{itemize}

%

%
\subsection{Implementation}

Our implementation of the incremental Gauss-Jordan xor-reasoning module uses a
dense matrix representation where one element in the matrix uses one bit of
memory. The xor-reasoning module maintains two such matrices. In the first
matrix the rows are consecutively in the memory, and in the second the columns
are consecutively in the memory. The first matrix allows efficient
implementation for row operations and the second matrix for efficient pivoting.
To detect xor-implied literals, each row is associated with a counter tracking
the number of unassigned variables.
When this counter is one (or zero), an xor-implied literal (or a potential
conflict) is available. Upon backtracking it suffices to restore the
counters tracking unassigned variables. Gauss-Jordan xor-reasoning module
is only used after unit propagation is saturated. To strengthen unit propagation over xor-constraints, explanations for xor-implied
literals are added as learned xor-constraints.

%
\subsection{Experimental Evaluation}

\begin{figure}[t]
  \centering
  \includegraphics[width=0.48\textwidth]{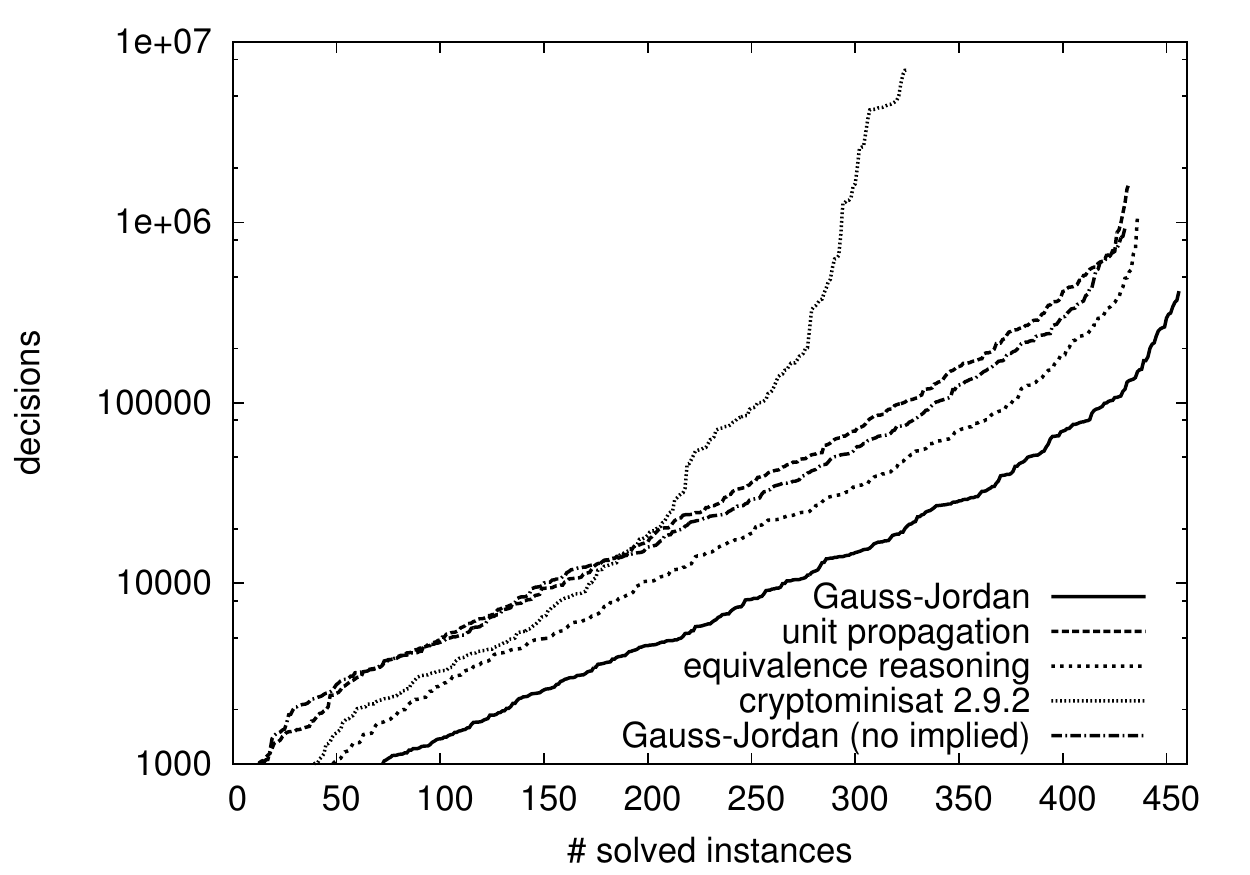} \\
  \includegraphics[width=0.48\textwidth]{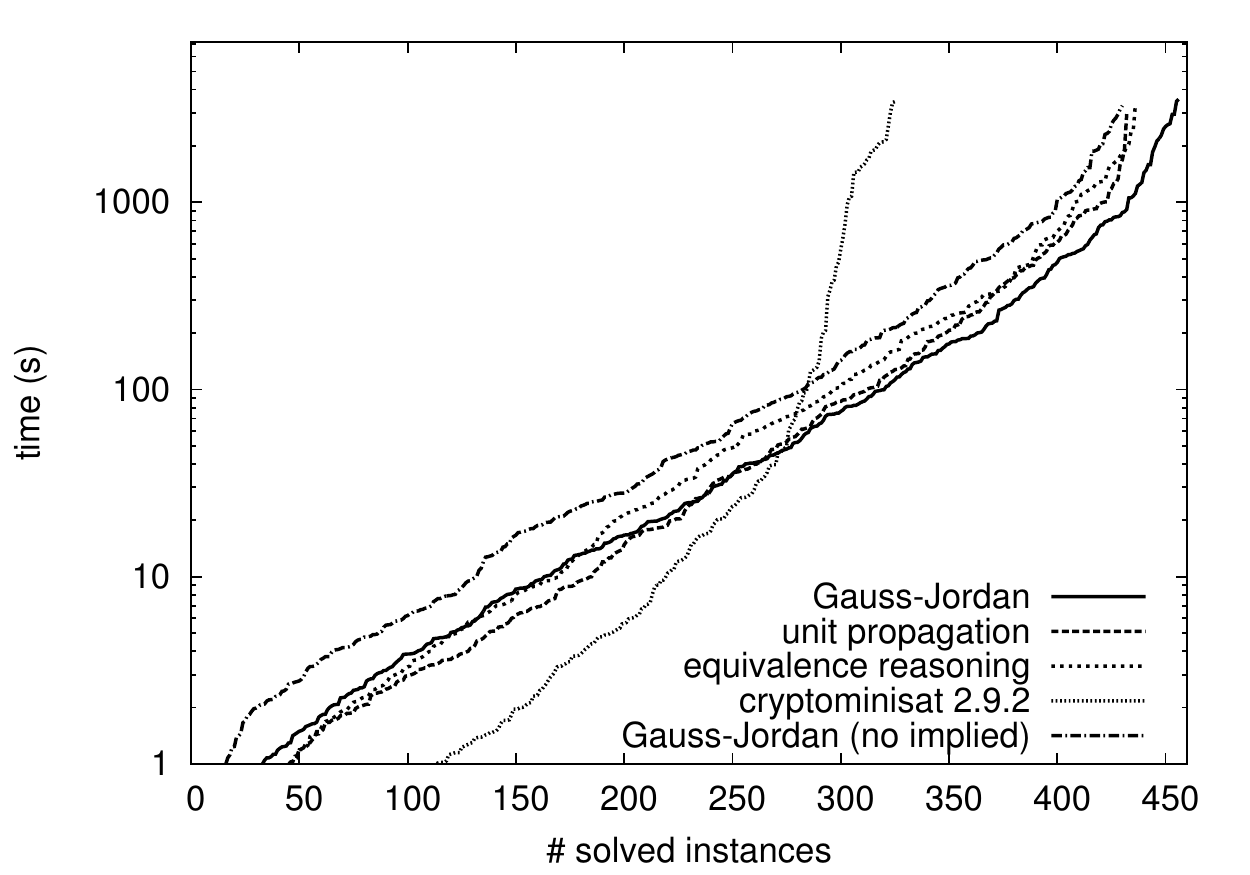}
  \caption{Comparison of three xor-reasoning modules (unit propagation, equivalence reasoning, Gauss-Jordan) and \STool{cryptominisat 2.9.2} on Trivium}
  \label{Fig:TriviumCactus}
\end{figure}

To evaluate the effect of incremental Gauss-Jordan elimination in the DPLL(XOR)
framework, we integrated three xor-reasoning modules with different
deduction engines (unit propagation, equivalence reasoning, Gauss-Jordan)
to \Tool{minisat 2.0 core}. In this experiment, we focus on the domain of logical
cryptanalysis by modeling a known-plaintext attack on stream cipher
Trivium. The task is to recover the full 80-bit key when the IV and a
number of cipher stream bits (8 to 16) are given. All instances are
satisfiable and it is likely that a number of keys produce the same given
prefix of the cipher stream. Figure~\ref{Fig:TriviumCactus} shows how unit
propagation, equivalence reasoning, incremental Gauss-Jordan and \Tool{cryptominisat 2.9.2} perform on these instances. The strength of the deduction engine is well
reflected in the results. The solver configuration relying only on unit
propagation requires the most decisions. Equivalence reasoning gives a significant reduction in the number of decisions and enables the solver to solve more instances. The solver configuration using incremental Gauss-Jordan solves the
highest number of instances and using fewest number of decisions. 
Considering the solving time, unit propagation can be implemented very
efficiently, so easier instances are solved fastest using plain unit
propagation. Equivalence reasoning incurs an additional computational overhead which causes it to perform slower than unit propagation despite the reduction in the number of decisions. Incremental Gauss-Jordan is computationally intensive but the reduction in the number of decisions is large enough to make it scale better for the harder instances.  To illustrate the effect of
        xor-implied literals deduced by Gauss-Jordan, a solver configuration using
        Gauss-Jordan only to detect conflicts and otherwise resorting to unit
        propagation is included in the comparison. Detecting conflicts as early
        as possible does not seem to help on this benchmark. The 
        lack of performance of \Tool{cryptominisat 2.9.2} is probably due to
        differences in restart policies or other heuristics. Gaussian
        elimination as implemented in \Tool{cryptominisat 2.9.2} using row echelon form does not seem to be very useful
        in this benchmark because on majority of the instances it does not
        detect conflicts earlier nor give any xor-implied literals.

\section{Exploiting Biconnected Components}

When using a dense representation for matrices
in the xor-reasoning modules based on Gauss or Gauss-Jordan elimination,
the worst-case memory use is $\BigOh(n e)$,
where $n$ is the number of variables 
and
$e$ the number of linearly independent xor-constraints in $\xorpart$.
Naturally,
when the xor-part $\xorpart$ can be decomposed into variable-disjoint sets of
xor-constraints (connected components of the constraint graph formally defined
below),
each such set can be handled by a separate xor-reasoning module
with smaller memory requirements.
When using a sparse matrix representation,
the memory usage does not improve with such a connected component decomposition.

We now give an improved decomposition technique that is based
on a new decomposition theorem stating that,
in order to guarantee full propagation,
it is enough to
(i) propagate only values through ``cut variables'',
and
(ii) have full propagation for the ``biconnected components''
between the cut variables.
Thus equivalences and more complicated relationships between variables
in different biconnected components do not have to be considered and
each component 
can be handled by a separate xor-reasoning module.
%

\newcommand{\VA}{V_\textup{a}}
\newcommand{\VB}{V_\textup{b}}

Formally,
given an xor-constraint conjunction $\xorpart$,
we define that a \emph{cut variable} is a variable $\Var \in \VarsOf{\xorpart}$
for which
there is a partition $(\VA,\VB)$ of xor-constraints in $\xorpart$
with ${\VarsOf{\VA} \cap \VarsOf{\VB}} = \Set{\Var}$;
such a partition $(\VA,\VB)$ is called an \emph{$\Var$-cut partition} of $\xorpart$.
The \emph{biconnected components} of $\xorpart$
are defined to be the equivalence classes
in the reflexive and transitive closure of the relation
$\Setdef{(\XC,\XCB)}{\textup{$\XC$ and $\XCB$ share a non-cut variable}}$
over the xor-constraints in $\xorpart$.
%
%
\newcommand{\Part}[1]{\mathcal{P}_{#1}}
\begin{example}
  \label{Ex:BiCo}
  Let $\xorpart = {(a \XX b \XX c\Equal\T)} \land {(b \XX d \XX e\Equal\T)} \land {(c \X e\Equal\T)} \land {(d \X e \XX f\Equal\F)} \land {(f \X g \X h\Equal\T)} \land (h \X i \X j\Equal\F) \land {(i \X j \X k\Equal\T)} \land {(f \X l \X m\Equal\T)} \land (l \X n \X o\Equal\F)$.
  %
  %
  The cut variables of $\xorpart$ are $f$, $h$ and $l$.
  Thus its five biconnected components are
  (i)
  $\{(a \X b \X c\Equal\T),(b \X d \X e\Equal\T),(c \X e\Equal\T),(d \X e \X f\Equal\F)\}$,
  (ii) $\Set{(f \XX g \XX h\Equal\T)}$,
  (iii) $\Set{(h \XX i \XX j\Equal\F),(i \XX j \XX k\Equal\T)}$,
  (iv) $\Set{(f \XX l \XX m\Equal\T)}$,
  and  
  (v) $\Set{(l \X n \X o\Equal\F)}$.
\end{example}

%

\newcommand{\CGraph}{G}
\newcommand{\CGNodes}{V}
\newcommand{\CGVNodes}{V_\textup{vars}}
\newcommand{\CGCNodes}{V_\textup{constrs}}
\newcommand{\CGEdges}{E}
\newcommand{\CGLab}{L}

\begin{figure}
  \centering
  \includegraphics[width=.95\columnwidth]{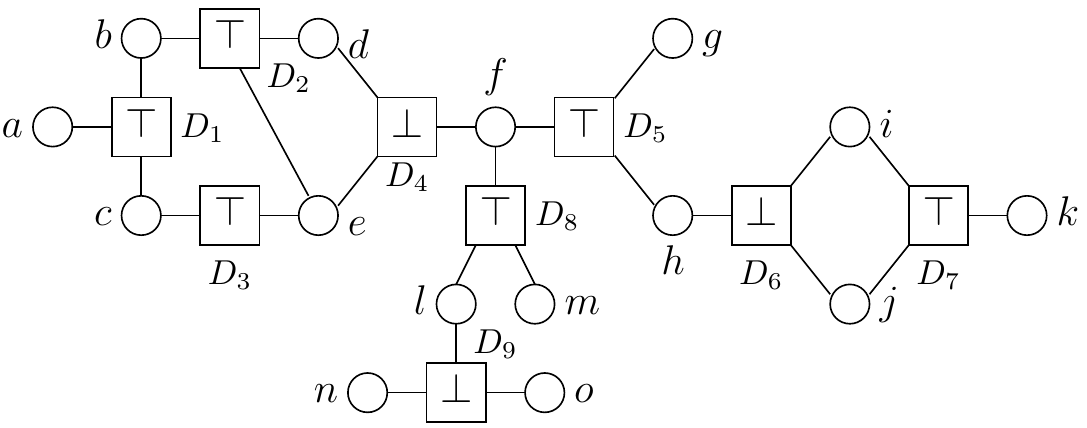}
  \caption{The constraint graph of the conjunction $\xorpart$ in Ex.~\ref{Ex:BiCo}}
  \label{Fig:CG}
\end{figure}

Cut variables and biconnected components
are probably best illustrated by means of constraint graphs.
Such graphs also give us a method for computing the cut variables,
and consequently also the biconnected components.
%
%
The \emph{constraint graph} of an xor-constraint conjunction $\xorpart$
is a labeled bipartite graph $\CGraph = \Tuple{\CGNodes, \CGEdges, \CGLab}$,
where
\begin{itemize}
\item
  the set of vertices $\CGNodes$
  is the disjoint union of
  (i)
    \emph{variable vertices} $\CGVNodes = \VarsOf{\xorpart}$
    which are graphically represented with circles,
    and
  (ii)
    \emph{constraint vertices} $\CGCNodes = \Setdef{\XC}{\text{$\XC$ is an xor-constraint in $\xorpart$}}$ drawn as rectangles,
\item
  $\CGEdges = {\Setdef{\Set{\Var,\XC}}{{\Var \in \CGVNodes} \land {\XC \in \CGCNodes} \land {\Var \in \VarsOf{\XC}}}}$
  are
  the edges connecting the variables
  and
  the xor-constraints in which they occur,
  and
\item
  $\CGLab$ labels each xor-constraint vertex $(\Var_1 \X ... \X \Var_k \Equal \parity{})$
  with the parity $\parity{}$.
\end{itemize}
As usual for graphs,
(i) a \emph{connected component} of constraint graph $\CGraph$
is a maximal connected subgraph of $\CGraph$,
(ii) a \emph{cut vertex} of $\CGraph$ is a vertex in it whose removal will break a connected component of $\CGraph$ into two or more connected components,
and
(iii) a \emph{biconnected component} of $\CGraph$ is a maximal biconnected subgraph (a graph is biconnected if it is connected and removing any vertex leaves the graph connected).
\begin{example}
  \label{Ex:BiCoGraph}
  The constraint graph of the conjunction $\xorpart$ in Ex.~\ref{Ex:BiCo}
  is shown in Fig.~\ref{Fig:CG}.
  The cut vertices of it are $\XCI1$, $\XCI4$, $f$, $\XCI5$, $h$, $\XCI6$, $\XCI7$, $\XCI8$, $l$, and $\XCI9$.
  Its biconnected components are the subgraphs induced by the vertex sets
  $\Set{a,\XCI1}$,
  $\Set{\XCI1,b,\XCI2,d,c,\XCI3,e,\XCI4}$,
  $\Set{\XCI4,f}$, and so on.
  Observe that the biconnected components are not vertex-disjoint.
\end{example}

We see that, due to the presense of the vertices for the xor-constraints,
the biconnected components of a constraint graph $\CGraph$ for $\xorpart$
do not directly correspond to the biconnected components of $\xorpart$.
However, the cut vertices of $\CGraph$,
when restricted to variable vertices,
correspond exactly to the cut variables of $\xorpart$.
Therefore, we have a linear time algorithm for computing the biconnected components of $\xorpart$:
\begin{enumerate}
\item
  Build (implicitly) the constraint graph $\CGraph$ for $\xorpart$.
\item
  Use an algorithm by Hopcroft and Tarjan~\cite{HopcroftTarjan:CACM1973}
  to compute the biconnected components of $\CGraph$ in linear time;
  as a byproduct, one gets all the cut vertices and thus the cut variables
  as well.
\item
  Build the biconnected components of $\xorpart$
  by putting two xor-constraints in the same component if they share a non-cut variable.
\end{enumerate}



%
\subsection{How to Exploit}

\newcommand{\xorpartA}{\xorpart^\textup{a}}
\newcommand{\xorpartB}{\xorpart^\textup{b}}

As biconnected components are connected to each other only through cut variables,
in the DPLL(XOR) framework we can actually handle them by separate xor-reasoning modules.
In this setting a value for a cut variable deduced by some xor-reasoning module
is communicated back to the CNF-part solver as an xor-implied literal,
and
the CNF-part solver then gives the value as an xor-assumption to the other
xor-reasoning modules.
Based on the following theorem,
we see that this kind of decomposition of $\xorpart$ preserves full propagation
in the following sense:
if the modules can provide full propagation for each of the components,
then full propagation is achieved for the whole xor-part $\xorpart$, too.
%
%
Basically the theorem states that only cut variable values,
not equivalences or more complex relationships,
have to be communicated between biconnected components.
For relating the theorem to biconnected components,
see the example after the theorem and
observe that if $(\VA,\VB)$ is an $\Var$-cut partition of $\xorpart$,
then $\VA$ and $\VB$ are (disjoint) unions of one or more biconnected
components of $\xorpart$.
%
%
%
\begin{theorem}\label{Thm:Decomposition}
  Let $(\VA,\VB)$ be an $\Var$-cut partition of $\xorpart$.
  Let $\xorpartA = \bigwedge_{\XC \in \VA}\XC$,
  $\xorpartB = \bigwedge_{\XC \in \VB}\XC$, and
  $\AL_1,...,\AL_k,\IL \in\LitsOf{\xorpart}$.
  Then it holds that:
  \begin{itemize}
  \item
    If $\xorpart \land {\AL_1 \land ... \land \AL_k}$ is unsatisfiable,
    then
    \begin{enumerate}
    \item
      $\xorpartA \land {\AL_1 \land ... \land \AL_k}$ or
      $\xorpartB \land {\AL_1 \land ... \land \AL_k}$ is unsatisfiable;
      or
    \item
      $\xorpartA \land {\AL_1 \land ... \land \AL_k} \Models (\Var \Equal \parity{\Var})$ and $\xorpartB \land {\AL_1 \land ... \land \AL_k} \Models (\Var \Equal \parity{\Var} \X \T)$ for some $\parity{\Var} \in \Set{\F,\T}$.
    \end{enumerate}
  \item
    If $\xorpart \land {\AL_1 \land ... \land \AL_k}$ is satisfiable
    and
    $\xorpart \land {\AL_1 \land ... \land \AL_k} \Models \IL$,
    then
    \begin{enumerate}
    \item
      $\xorpartA \land {\AL_1 \land ... \land \AL_k} \Models \IL$ or
      $\xorpartB \land {\AL_1 \land ... \land \AL_k} \Models \IL$; or
    \item
      $\xorpartA \land {\AL_1 \land ... \land \AL_k} \Models (\Var \Equal \parity{\Var})$ and $\xorpartB \land {\AL_1 \land ... \land \AL_k \land (\Var \Equal \parity{\Var})} \Models \IL$;
      or
    \item
      $\xorpartB \land {\AL_1 \land ... \land \AL_k} \Models (\Var \Equal \parity{\Var})$ and $\xorpartA \land {\AL_1 \land ... \land \AL_k \land (\Var \Equal \parity{\Var})} \Models \IL$.
    \end{enumerate}
  \end{itemize}
\end{theorem}

\begin{example}
  Take again the conjunction $\xorpart$ in Ex.~\ref{Ex:BiCo},
  illustrated in Fig.~\ref{Fig:CG}.
  Assume the xor-assumptions $b$, ${\neg g}$, and $o$;
  now $\xorpart \land b \land {\neg g} \land o \Models {\neg k}$.
  We can deduce this in a biconnected component-wise manner as follows.
  First, consider the $f$-cut partition
  $(\Set{\XCI1,...,\XCI4},\Set{\XCI5,...,\XCI9})$.
  Now $\XCI1 \land ... \land \XCI4 \land b \Models {\neg f}$
  and
  $\XCI5 \land ... \land \XCI9 \land {\neg g} \land o \land {\neg f} \Models {\neg k}$.
  For $\XCI5 \land ... \land \XCI9 \land {\neg g} \land o \land {\neg f} \Models {\neg k}$
  we apply the theorem again
  by considering the $f$-cut partition
  $(\Set{\XCI8,\XCI9},\Set{\XCI5,\XCI6,\XCI7})$
  of $\XCI5 \land ... \land \XCI9$:
  now $\XCI5 \land \XCI6 \land \XCI7 \land {\neg g} \land {\neg f} \Models {\neg k}$
  and
  thus the biconnected components $\Set{\XCI8}$ and $\Set{\XCI9}$ are not needed in the derivation.
  For $\XCI5 \land \XCI6\land \XCI7 \land {\neg g} \land {\neg f} \Models {\neg k}$,
  we apply the theorem again
  by considering the $h$-cut partition
  $(\Set{\XCI5},\Set{\XCI6,\XCI7})$:
  $\XCI5 \land {\neg g} \land {\neg f} \Models h$
  and
  $\XCI6 \land \XCI7 \land h \Models {\neg k}$.
  Thus we can derive $\neg k$ from $\xorpart \land b \land {\neg g}$
  in a component-by-component fashion.
\end{example}

We observe the following:
some biconnected components can be singleton sets.
For such components we can provide full propagation easily by the basic
unit propagation.
These singleton components originate from ``tree-like'' parts of $\xorpart$:
the trees can be ``outermost'' (constraints $\XCI8$ and $\XCI9$ in Fig.~\ref{Fig:CG})
or between two non-tree-like components ($\XCI5$ in Fig.~\ref{Fig:CG}).
Thus our new result in a sense subsumes one in \cite{LJN:CP2012},
where we suggested clausification of ``outermost'' tree-like parts.

%
\subsection{Experimental Evaluation}

\begin{figure}
  \centering
    \includegraphics[width=\columnwidth]{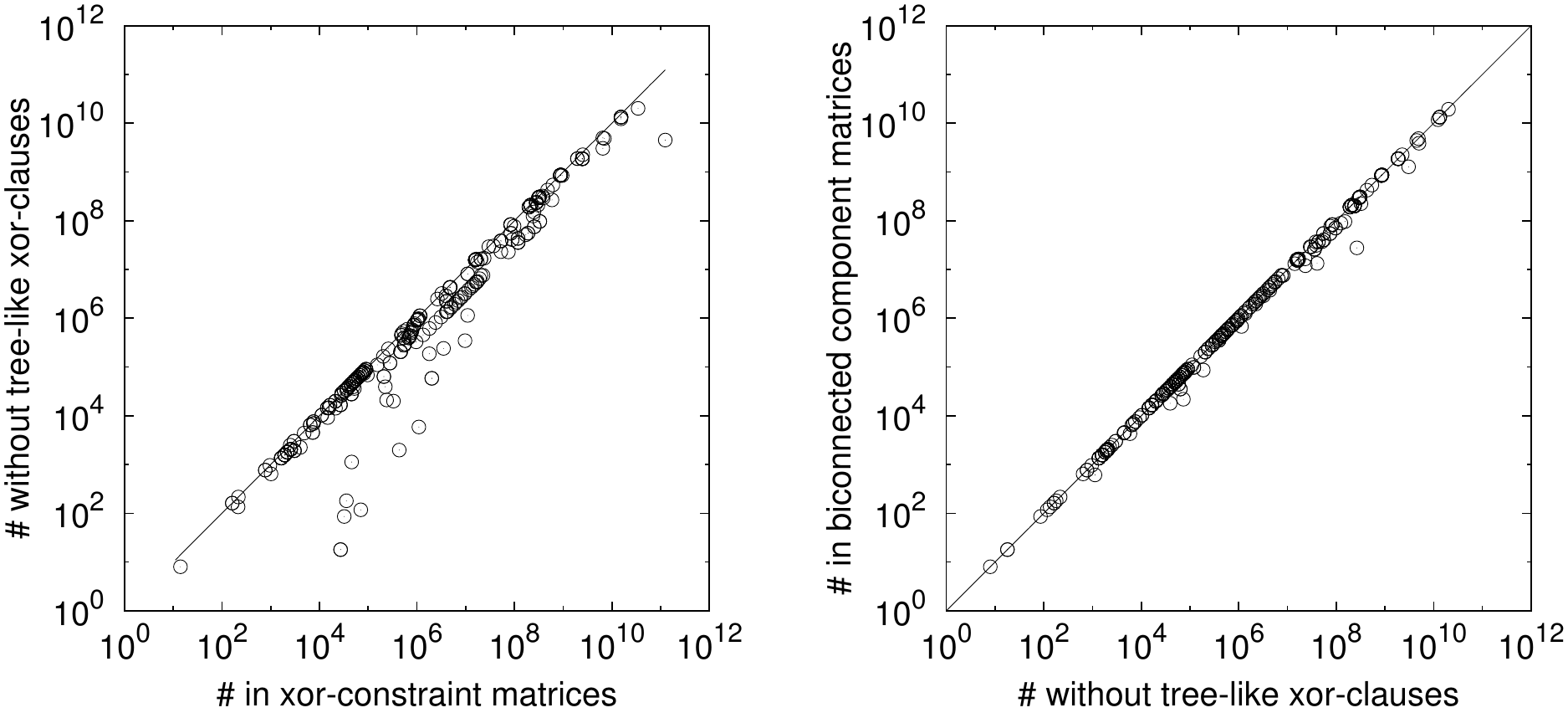} 
  \caption{Reduction in memory usage for dense matrix representation when (i) tree-like xor-constraints are removed (ii) only biconnected components are counted in SAT 2005-2011 competition instances. In the latter case, although the difference seems negligible in logarithmic scale, the memory consumption is reduced by additional 13.5\% on average in 110 instances having multiple biconnected components.}
  \label{Fig:MatrixSizes}
\end{figure}

To evaluate the relevance of detecting biconnected components, we studied
the benchmark instances in ``crafted'' and ``industrial/application''
categories of the SAT Competitions 2005, 2007, and 2009 as well as all the
instances in the SAT Competition 2011 (available at
\url{http://www.satcompetition.org/}). To get rid of some ``trivial'' xor-constraints,
we eliminated unary clauses and binary xor-constraints from each instance
by unit propagation and substitution, respectively.
After this easy preprocessing, 474 instances (with some duplicates due to
overlap in the competitions) having xor-constraints remained.
We first examine how the memory usage can be improved by removing (i) tree-like
xor-constraints and (ii) storing each biconnected component in a separate matrix.
Figure~\ref{Fig:MatrixSizes} shows the reduction in memory usage
when using dense matrix representation to store the xor-constraints. As already
reported in~\cite{LJN:CP2012},
a significant proportion of xor-constraints in these competition
instances are tree-like and performing additional reasoning beyond
unit propagation cannot be used to detect more implied literals. Removing these
tree-like xor-constraints from Gauss-Jordan matrices reduces the memory usage
greatly. An additional reduction in memory usage is obtained by storing each
biconnected component in a separate matrix. 

\begin{figure}
  \centering
  \includegraphics[width=0.45\textwidth]{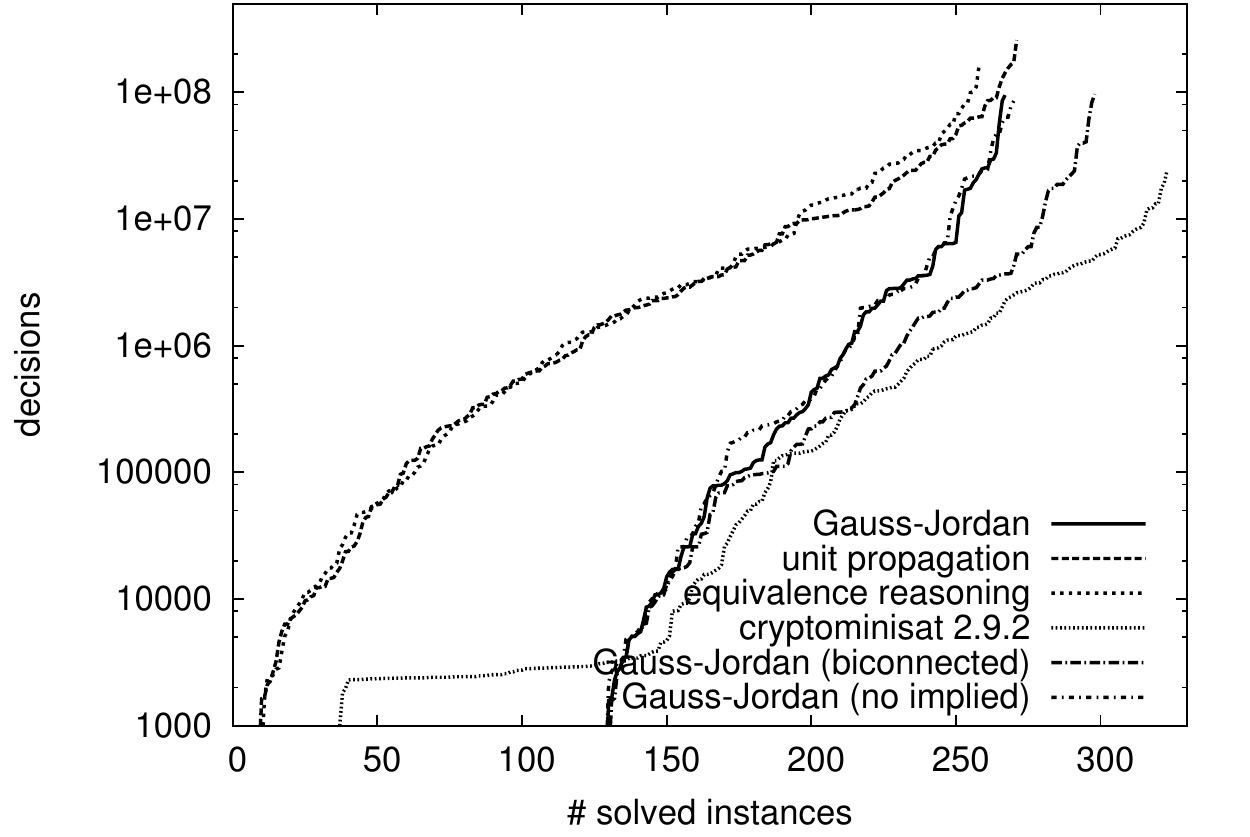} 
  \\
  \includegraphics[width=0.45\textwidth]{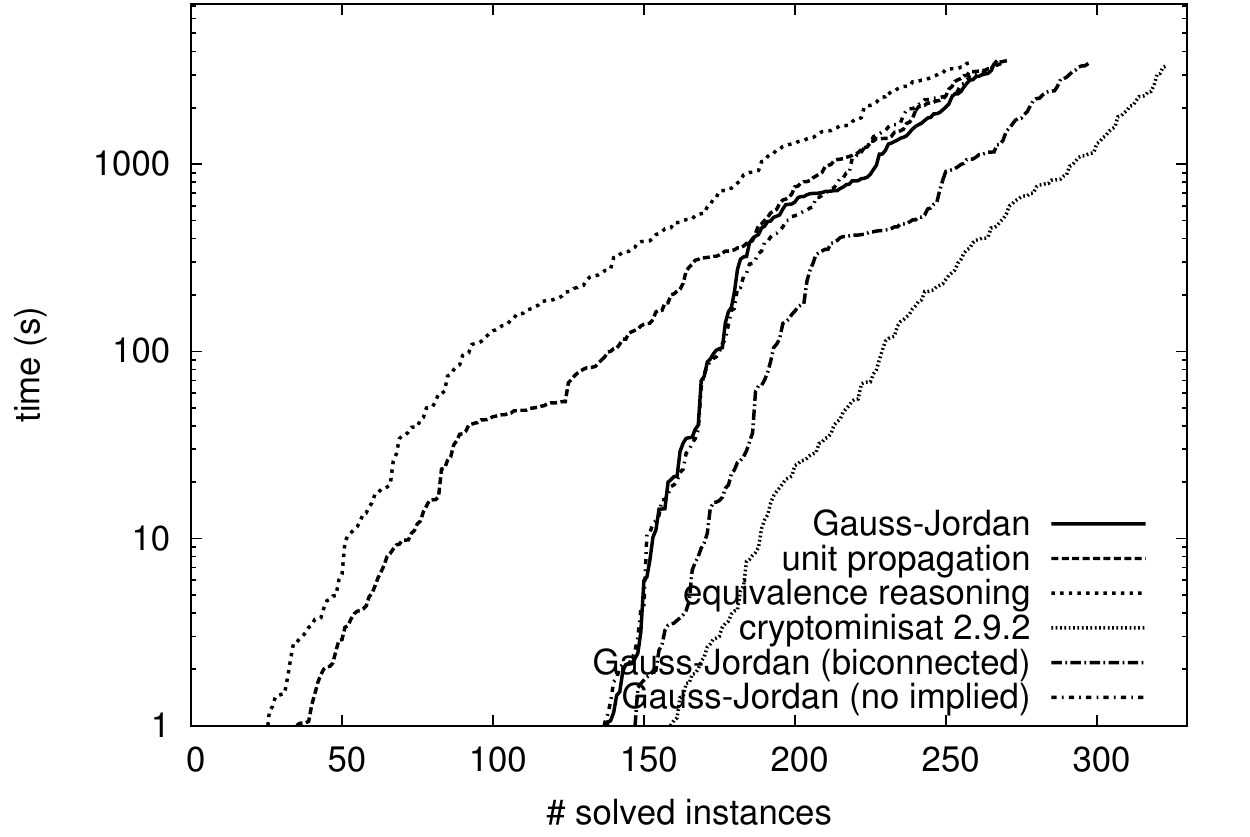} 
  \caption{Number of SAT 2005-2011 competition instances solved w.r.t.\ decisions and time}
  \label{Fig:SatDecisionsTimeCactus}
\end{figure}

We ran \Tool{minisat 2.0 core} augmented with four different xor-reasoning modules (unit
        propagation, equivalence reasoning, Gauss-Jordan, and a variant of
        Gauss-Jordan exploiting biconnected components) and \Tool{cryptominisat 2.9.2}
on these instances. Figure~\ref{Fig:SatDecisionsTimeCactus} shows the number of
instances solved with respect to the number of heuristic decisions. Unit
propagation and equivalence reasoning perform similarly on these instances.
Incremental Gauss-Jordan solves a substantial number of the instances almost
instantly and also manages to solve more instances in total. The solver
\Tool{cryptominisat 2.9.2} performs very well on these instances.
Figure~\ref{Fig:SatDecisionsTimeCactus} also
shows the number of instances solved with respect to time. Since equivalence reasoning does not reduce the number of
decisions, the computational overhead is reflected in the slowest solving time.
Incremental Gauss-Jordan is computationally more intensive but complete parity
reasoning pays off on these instances leading to fastest solving compared to
our other xor-reasoning modules. Omitting tree-like xor-constraints from
Gauss-Jordan matrices and splitting biconnected components into separate
matrices offers a significant reduction in the solving time without sacrificing
completeness of reasoning.  To illustrate the effect of implied literals
deduced by Gauss-Jordan, we also ran a solver using Gauss-Jordan
only to detect conflicts and otherwise resorting to unit propagation. More
instances are solved and faster when all implied literals are deduced.

\begin{figure}
    \includegraphics[width=0.48\textwidth]{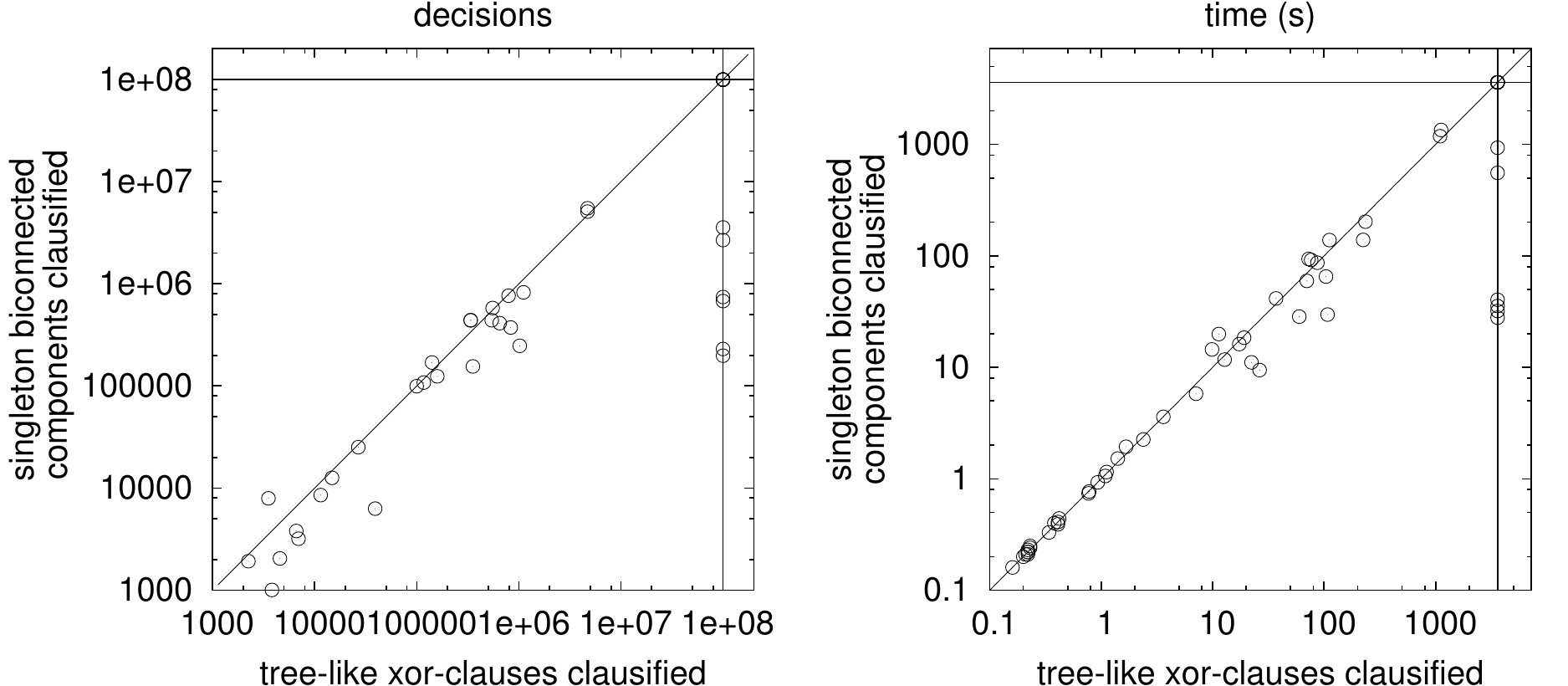} 
  \caption{Effect in decisions and solving time for \STool{cryptominisat} when singleton biconnected components in SAT competition instances are translated to CNF}
  \label{Fig:NonCyclicClausifiedCryptominisat}
\end{figure}

Biconnected components may be exploited even without modifying the
solver. The solver \Tool{cryptominisat} accepts a mixture of clauses and
xor-constraints as its input. When Gaussian elimination is used, the solver stores
each connected component in a separate matrix. By translating each singleton
biconnected component into CNF, some non-trivial biconnected components may
become connected components and are then placed into separate matrices
improving memory usage. We considered the 110 SAT competition
instances with multiple biconnected components and found 60 instances where
some biconnected components could be separated by translating singleton
biconnected components to CNF.
Figure~\ref{Fig:NonCyclicClausifiedCryptominisat} shows the effect of the
translation in the number of decisions and solving time.
The solver \Tool{cryptominisat 2.9.2} solves 44 of the unmodified instances. After the
translation, \Tool{cryptominisat 2.9.2} is able to solve 50 instances and slightly
faster.

\section{Eliminating xor-internal variables}

A cnf-xor formula $\orpart \wedge \xorpart$ may have {\it xor-internal}
variables occurring only in $\xorpart$. As suggested in~\cite{LJN:ECAI2010},
such variables can be eliminated from the formula by substituting
them with their ``definitions''; e.g. if $ x_1 \oplus x_2 \oplus x_3 \Equal \top
$ is an xor-constraint where $x_1$ is an xor-internal variable, then
remove the parity constraint and replace every occurrence of $x_1$ in
all the other parity constraints by $ x_2 \oplus x_3 \oplus \top $.
When using dense matrix representation, the matrices can be made 
more compact by eliminating xor-internal
variables. For instance, one of our Trivium benchmark instances has
5900 xor-internal variables out of 11484 variables and 8590 parity
constraints in two connected components. The total number of elements
in the matrices is $55 \times 10^6$ elements. By eliminating all
xor-internal variables this can be reduced to $8 \times 10^6$
elements. The instance has three biconnected components (as all of
our Trivium instances) and storing them in separate matrices
requires $ 33 \times 10^6$ elements in total. But, if a cut
variable connecting the biconnected components is xor-internal, it is
eliminated and the two biconnected components are merged into one
bigger biconnected component. To preserve biconnected
components, only the variables occurring in a single biconnected
component and not in the CNF-part should be eliminated. There are 5906
such variables in the instances and after the elimination the total
number of elements in three matrices is $5 \times 10^6$.
Figure~\ref{Fig:TriviumElimCactus} shows the effect of eliminating
such variables in our Trivium instances. Unit propagation benefits from
elimination of xor-internal variables. Fewer watched
literals (variables) are needed for longer xor-constraints to detect when an implied
literal can be deduced. The solver configuration using incremental
Gauss-Jordan elimination manages to solve all of our benchmark instances
with reduced solving time.
\iffalse
\begin{figure*}
  \centering
    \includegraphics[width=0.45\textwidth]{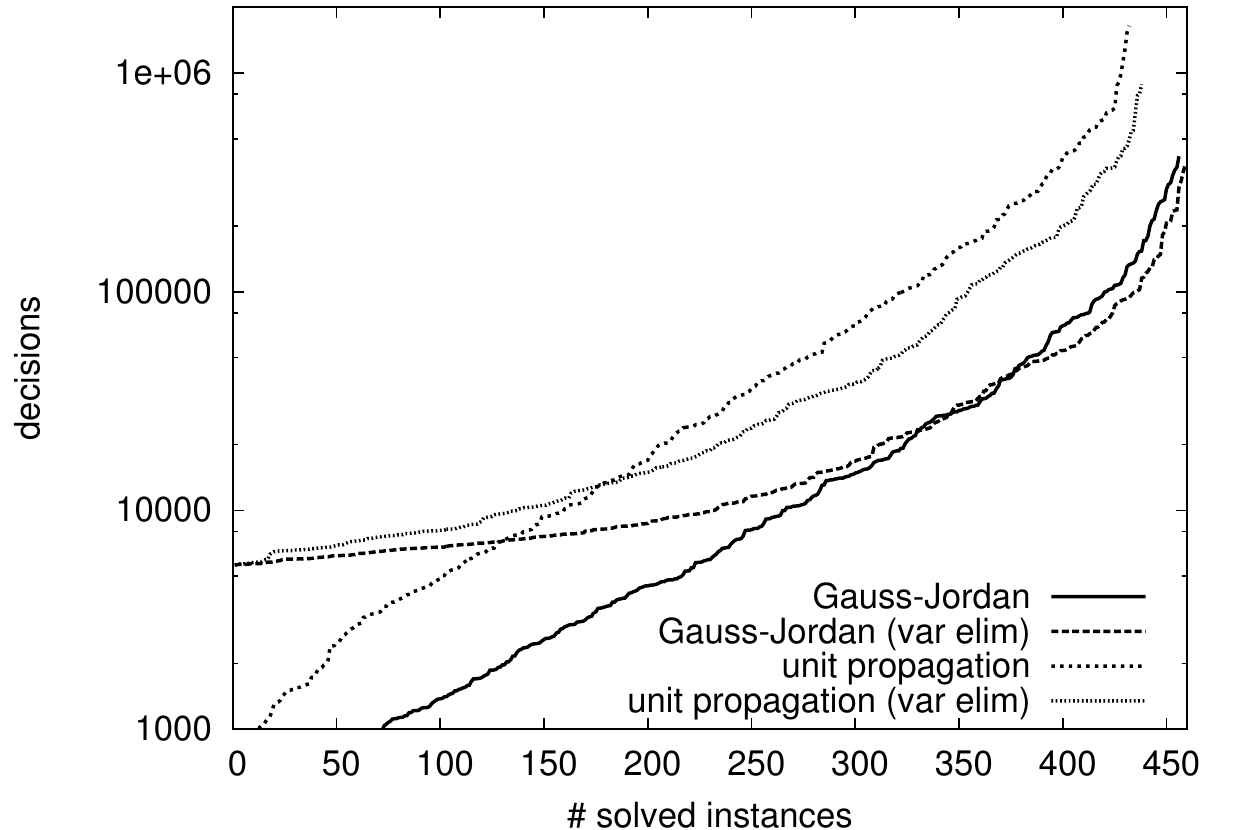} \qquad
    \includegraphics[width=0.45\textwidth]{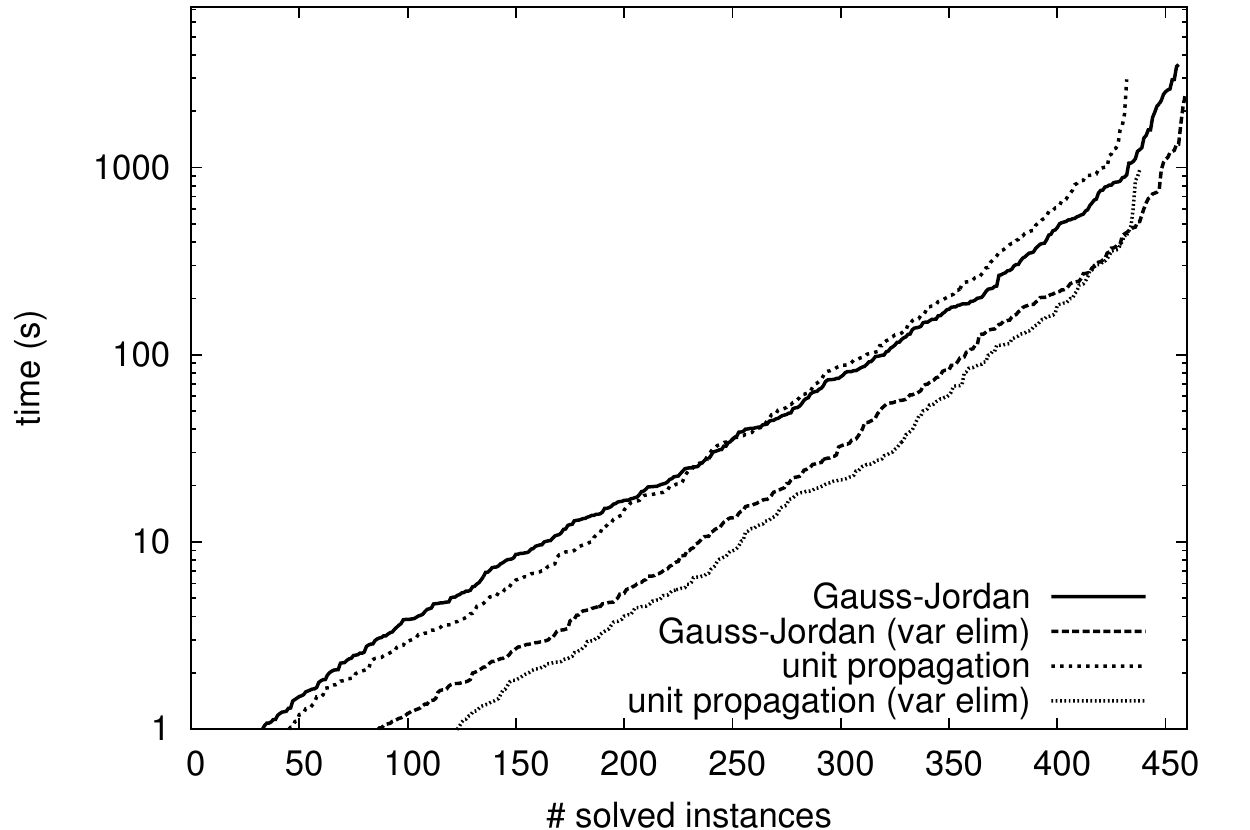} 
  \caption{Effect of eliminating xor-internal variables while preserving biconnected components in the number of decisions and solving time on Trivium}
  \label{Fig:TriviumElimCactus}
\end{figure*}
\else
\begin{figure}
  \centering
    \includegraphics[width=0.45\textwidth]{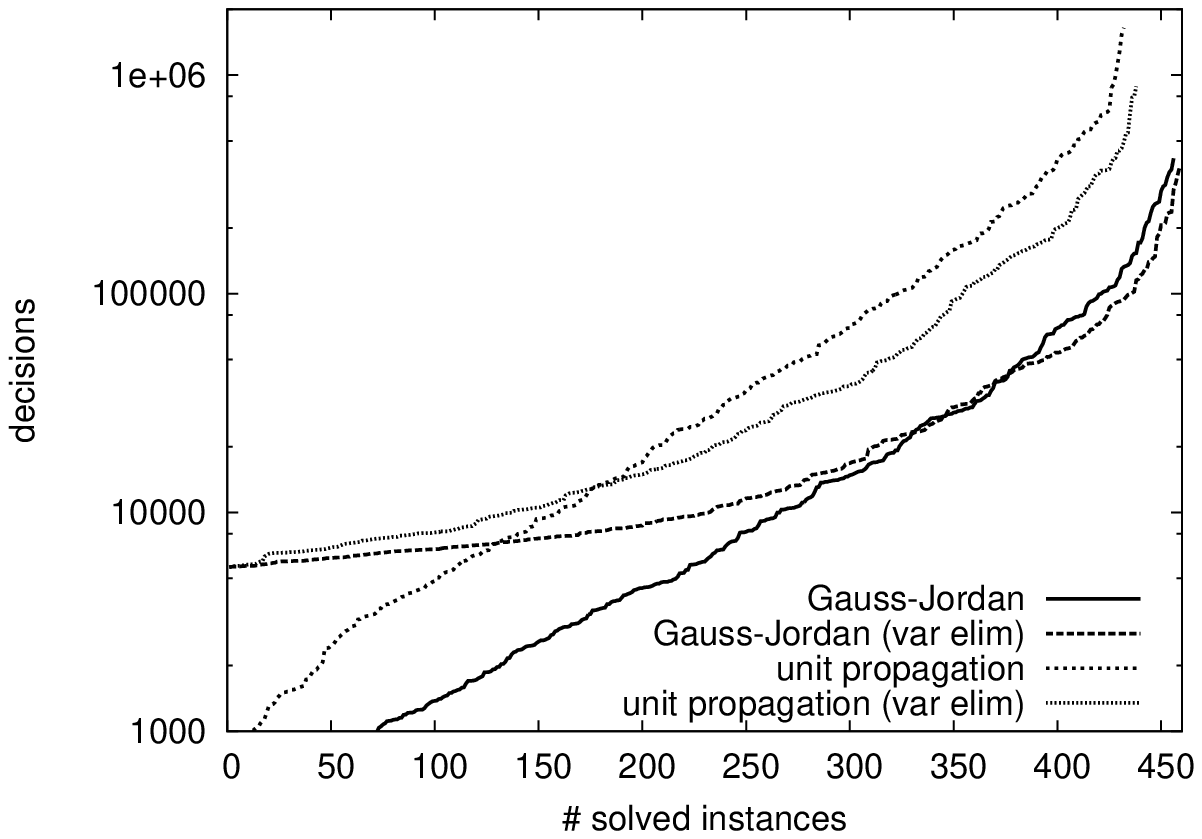} \\
    \includegraphics[width=0.45\textwidth]{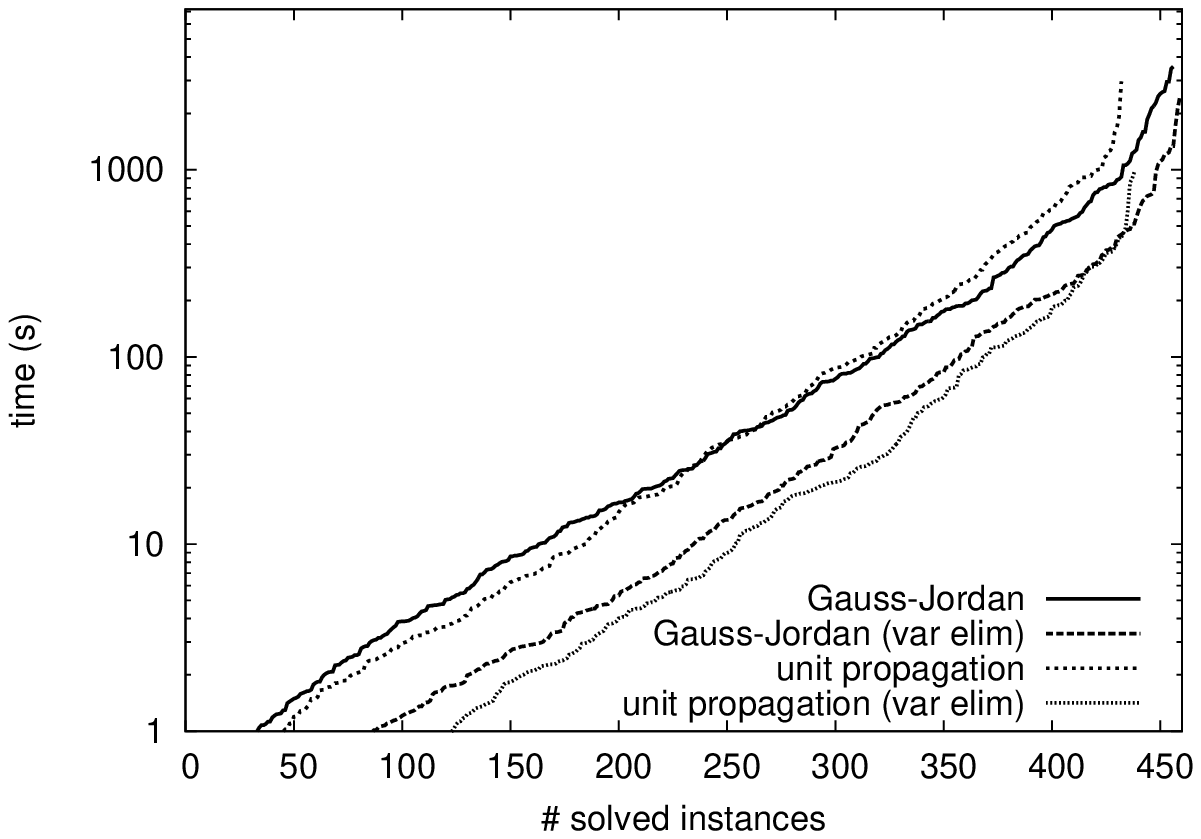} 
  \caption{Effect of eliminating xor-internal variables while preserving biconnected components in the number of decisions and solving time on Trivium}
  \label{Fig:TriviumElimCactus}
\end{figure}
\fi

\section*{Acknowledgment}

This work has been financially supported by the Academy  of Finland under the Finnish Centre of Excellence in Computational Inference (COIN).



\bibliographystyle{IEEEtran}
\bibliography{fullpaper}

\begin{thebibliography}{10}
\providecommand{\url}[1]{#1}
\csname url@samestyle\endcsname
\providecommand{\newblock}{\relax}
\providecommand{\bibinfo}[2]{#2}
\providecommand{\BIBentrySTDinterwordspacing}{\spaceskip=0pt\relax}
\providecommand{\BIBentryALTinterwordstretchfactor}{4}
\providecommand{\BIBentryALTinterwordspacing}{\spaceskip=\fontdimen2\font plus
\BIBentryALTinterwordstretchfactor\fontdimen3\font minus
  \fontdimen4\font\relax}
\providecommand{\BIBforeignlanguage}[2]{{%
\expandafter\ifx\csname l@#1\endcsname\relax
\typeout{** WARNING: IEEEtran.bst: No hyphenation pattern has been}%
\typeout{** loaded for the language `#1'. Using the pattern for}%
\typeout{** the default language instead.}%
\else
\language=\csname l@#1\endcsname
\fi
#2}}
\providecommand{\BIBdecl}{\relax}
\BIBdecl

\bibitem{Handbook:CDCL}
J.~Marques-Silva, I.~Lynce, and S.~Malik, ``Conflict-driven clause learning
  {SAT} solvers,'' in \emph{Handbook of Satisfiability}.\hskip 1em plus 0.5em
  minus 0.4em\relax IOS Press, 2009.

\bibitem{Urquhart:JACM1987}
A.~Urquhart, ``Hard examples for resolution,'' \emph{Journal of the ACM},
  vol.~34, no.~1, pp. 209--219, 1987.

\bibitem{Li:AAAI2000}
C.~M. Li, ``Integrating equivalency reasoning into {Davis-Putnam} procedure,''
  in \emph{Proc.~AAAI/IAAI 2000}.\hskip 1em plus 0.5em minus 0.4em\relax AAAI
  Press, 2000, pp. 291--296.

\bibitem{Li:IPL2000}
------, ``Equivalency reasoning to solve a class of hard {SAT} problems,''
  \emph{Information Processing Letters}, vol.~76, no. 1--2, pp. 75--81, 2000.

\bibitem{BaumgartnerMassacci:CL2000}
P.~Baumgartner and F.~Massacci, ``The taming of the {(X)OR},'' in
  \emph{Proc.~CL~2000}, ser. LNCS, vol. 1861.\hskip 1em plus 0.5em minus
  0.4em\relax Springer, 2000, pp. 508--522.

\bibitem{Li:DAM2003}
C.~M. Li, ``Equivalent literal propagation in the {DLL} procedure,''
  \emph{Discrete Applied Mathematics}, vol. 130, no.~2, pp. 251--276, 2003.

\bibitem{HeuleMaaren:SAT2004}
M.~Heule and H.~van Maaren, ``Aligning {CNF}- and equivalence-reasoning,'' in
  \emph{Proc.\ SAT 2004}, ser. LNCS, vol. 3542.\hskip 1em plus 0.5em minus
  0.4em\relax Springer, 2004, pp. 145--156.

\bibitem{HeuleEtAl:SAT2004}
M.~Heule, M.~Dufour, J.~van Zwieten, and H.~van Maaren, ``March\_eq:
  Implementing additional reasoning into an efficient look-ahead {SAT}
  solver,'' in \emph{Proc.\ SAT 2004}, ser. LNCS, vol. 3542.\hskip 1em plus
  0.5em minus 0.4em\relax Springer, 2004, pp. 345--359.

\bibitem{Chen:SAT2009}
J.~Chen, ``Building a hybrid {SAT} solver via conflict-driven, look-ahead and
  {XOR} reasoning techniques,'' in \emph{Proc.\ SAT 2009}, ser. LNCS, vol.
  5584.\hskip 1em plus 0.5em minus 0.4em\relax Springer, 2009, pp. 298--311.

\bibitem{SoosEtAl:SAT2009}
M.~Soos, K.~Nohl, and C.~Castelluccia, ``Extending {SAT} solvers to
  cryptographic problems,'' in \emph{Proc.\ SAT 2009}, ser. LNCS, vol.
  5584.\hskip 1em plus 0.5em minus 0.4em\relax Springer, 2009, pp. 244--257.

\bibitem{LJN:ECAI2010}
T.~Laitinen, T.~Junttila, and I.~Niemel{\"a}, ``Extending clause learning
  {DPLL} with parity reasoning,'' in \emph{Proc.\ ECAI 2010}.\hskip 1em plus
  0.5em minus 0.4em\relax IOS Press, 2010, pp. 21--26.

\bibitem{Soos}
M.~Soos, ``Enhanced gaussian elimination in {DPLL}-based {SAT} solvers,'' in
  \emph{Pragmatics of SAT}, Edinburgh, Scotland, GB, July 2010, pp. 1--1.

\bibitem{LJN:ICTAI2011}
T.~Laitinen, T.~Junttila, and I.~Niemel{\"a}, ``Equivalence class based parity
  reasoning with {DPLL(XOR)},'' in \emph{Proc.\ ICTAI 2011}.\hskip 1em plus
  0.5em minus 0.4em\relax IEEE, 2011, pp. 649--658.

\bibitem{LJN:SAT2012}
------, ``Conflict-driven {XOR}-clause learning,'' in \emph{Proc.\ SAT 2012},
  ser. LNCS, vol. 7317.\hskip 1em plus 0.5em minus 0.4em\relax Springer, 2012,
  pp. 383--396.

\bibitem{HanJiang:CAV2012}
C.-S. Han and J.-H.~R. Jiang, ``When boolean satisfiability meets gaussian
  elimination in a simplex way,'' in \emph{Proc.\ CAV 2012}, 2012, to appear.

\bibitem{NieuwenhuisEtAl:JACM06}
R.~Nieuwenhuis, A.~Oliveras, and C.~Tinelli, ``Solving {SAT} and {SAT} modulo
  theories: From an abstract {D}avis-{P}utnam-{L}ogemann-{L}oveland procedure
  to {DPLL(T)},'' \emph{Journal of the ACM}, vol.~53, no.~6, pp. 937--977,
  2006.

\bibitem{Handbook:SMT}
C.~Barrett, R.~Sebastiani, S.~A. Seshia, and C.~Tinelli, ``Satisfiability
  modulo theories,'' in \emph{Handbook of Satisfiability}.\hskip 1em plus 0.5em
  minus 0.4em\relax IOS Press, 2009.

\bibitem{DutertreMoura:CAV2006}
B.~Dutertre and L.~M. de~Moura, ``A fast linear-arithmetic solver for
  {DPLL(T)},'' in \emph{CAV}, ser. LNCS, vol. 4144.\hskip 1em plus 0.5em minus
  0.4em\relax Springer, 2006, pp. 81--94.

\bibitem{HopcroftTarjan:CACM1973}
J.~E. Hopcroft and R.~E. Tarjan, ``Efficient algorithms for graph manipulation
  [h] (algorithm 447),'' \emph{Communications of the ACM}, vol.~16, no.~6, pp.
  372--378, 1973.

\bibitem{LJN:CP2012}
T.~Laitinen, T.~Junttila, and I.~Niemel\"a, ``Classifying and propagating
  parity constraints,'' 2012, accepted for publication in CP 2012.

\end{thebibliography}

\newpage
\appendix
\section{Proofs}

\newenvironment{relemma}[1]{\renewcommand{\thetheorem}{#1}\begin{lemma}}{\end{lemma}}
\newenvironment{retheorem}[1]{\renewcommand{\thetheorem}{#1}\begin{theorem}}{\end{theorem}}

\newcommand{\LinComb}{+}
\newcommand{\BigLinComb}{\sum}


In this appendix, we provide proofs for the Lemmas and Theorems in the paper.
Before the actual proofs,
we provide some auxiliary results.

For two xor-constraints
$\XC = (x_1 \X ... \X x_k \Equal p)$
and
$\XCB = (y_1 \X ... \X y_l \Equal q)$,
we define their linear combination xor-constraint by
$\XC \LinComb \XCB =
 (x_1 \X ... \X x_k \X y_1 \X ... \X y_l \Equal p \X q)$.
Some fundamental, easy to verify properties are
$\XC \LinComb \XC \LinComb \XCB = \XCB$,
${\XC \land \XCB} \Models {\XC \LinComb \XCB}$,
${\XC \land \XCB} \Models {\XC \land (\XC \LinComb \XCB)}$,
and
${\XC \land (\XC \LinComb \XCB)} \Models {\XC \land \XCB}$.
Furthermore,
the logical consequence xor-constraints of a conjunction $\xorpart$
are exactly those that are linear combinations of the xor-constraints in $\xorpart$:
\begin{lemma}\label{Lemma:LinearCombs}
  Let $\psi$ be a conjunction of xor-constraints.
  Now $\psi$ is unsatisfiable if and only if
  there is a subset $S$ of xor-constraints in $\psi$ such that
  $\BigLinComb_{\XC \in S} \XC = (\F \Equal \T)$.
  If $\psi$ is satisfiable and $\XCB$ is an xor-constraint,
  then $\psi \Models \XCB$ 
  if and only if
  there is a subset $S$ of xor-constraints in $\psi$ such that
  $\BigLinComb_{\XC \in S} \XC = \XCB$.
\end{lemma}
\iftrue 
\begin{IEEEproof}
  There are two cases to consider.
  \begin{itemize}
  \item
  Case I: $\psi$ is unsatisfiable.

  If there is a subset $S$ of xor-constraints in $\psi$ such that
  $\BigLinComb_{\XC \in S} \XC = (\F \Equal \T)$,
  then,
  by iteratively applying ${\XCI1 \land \XCI2} \Models {\XCI1 \LinComb \XCI2}$,
  we have $\bigwedge_{\XC \in S} \XC \Models \BigLinComb_{\XC \in S} \XC$,
  i.e.~$\BigLinComb_{\XC \in S} \XC \Models (\F \Equal \T)$,
  and
  thus $\psi$ is unsatisfiable.

  For the other direction, assume that $\psi$ is unsatisfiable.
  Represent the conjunction $\psi$ as a system of linear equations
  modulo two in matrix form.
  Gaussian elimination must result in an equation $0 \equiv 1 \mod 2$
  in some row $r$ of the matrix.
  The row $r$ is a linear combination of some original rows
  $r_1, \dots, r_n$.
  Each original row $r_i$ 
  corresponds to a distinct xor-constraint $C(r_i)$ in $\psi$.
  Thus, $S = \set{C(r_1), \dots, C(r_n)} \subseteq \psi$
  is a subset of xor-constraints in $\psi$ such that
  $\BigLinComb_{\XC \in S} \XC = (\F \Equal \T)$.

  \item
  Case II: $\psi$ is satisfiable.

  If there is a subset $S$ of xor-constraints in $\psi$ such that
  $\BigLinComb_{\XC \in S} \XC = \XCB$,
  then,
  by iteratively applying ${\XCI1 \land \XCI2} \Models {\XCI1 \LinComb \XCI2}$,
  we have $\bigwedge_{\XC \in S} \XC \Models \BigLinComb_{\XC \in S} \XC$
  and
  thus $\bigwedge_{\XC \in S} \XC \Models \XCB$ and $\psi \Models \XCB$.


  Assume that $\psi \Models \XCB$.
  %
  We have $\emptyset \neq \VarsOf{\XCB} \subseteq \VarsOf{\psi}$.
  Create a (reduced row echelon form) tableau $\Eqs$ for $\psi$
  with the following property holding for each equation $\Eq$:
  if $\Eq$ has a non-basic variable occurring in $\XCB$,
  then the basic variable of $\Eq$ also occurs in $\XCB$.
  Such a tableau can be obtained by applying the $\operatorname{swap}$
  operator at most $\Card{\VarsOf{\XCB}}$ times to a tableau for $\psi$.
  By construction,
  each equation $\Eq$ of form ${x \Def x_1 \X ... \X x_k \X \parity{}}$ in $\Eqs$
  corresponds to a linear combination
  $C_\Eq = (x \X x_1 \X ... \X x_k \Equal \parity{})$
  of a subset
  $S_\Eq$ of xor-constraints in $\psi$.
  Consider the linear combination
  $\XCB' = \BigLinComb_{\Eq \in \Eqs \land {\VarsOf{\Eq} \cap \VarsOf{\XCB} \neq \emptyset}} C_\Eq$
  of equations in $\Eqs$ having at least one common variable with $\XCB$.
  It holds that $\psi \Models \XCB'$.
  As $\psi \Models \XCB$ and $\psi \Models \XCB'$,
  it also holds that $\psi \Models \XCB \land \XCB'$
  and thus $\psi \Models \XCB \LinComb \XCB'$.
  We have three cases to consider:
  \begin{itemize}
  \item
    Case A: $\XCB \LinComb \XCB' = (\F \Equal \T)$.
    This is not possible as $\psi$ would be unsatisfiable.
  \item
    Case B: $\XCB \LinComb \XCB' = (\F \Equal \F)$.
    Now $\XCB'$ is equal to $\XCB$.
    Thus there is a subset $S$ of xor-constraints in $\psi$
    such that $\BigLinComb_{\XC \in S} \XC = \XCB$,
    namely the ones that appear an odd number of times in
    $\bigcup_{\Eq \in \Eqs \land {\VarsOf{\Eq} \cap \VarsOf{\XCB} \neq \emptyset}} S_\Eq$
    (whose linear combination $\XCB'$ is).
  \item
    Case C:
    $\XCB \LinComb \XCB'$ is
    $y_1 \X ... \X y_k \Equal \parity{}$ with $k \ge 1$.
    All the variables $y_1,...,y_k$ must be non-basic variables in $\Eqs$
    because (i) all the basic variables of $\Eqs$ occurring in $\XCB$ also occur in $\XCB'$, and
    (ii) the basic variables of $\Eqs$ not occurring in $\XCB$ are not included in $\XCB'$ either.
    But because $y_1,...,y_k$ are non-basic variables,
    we can build the following satisfying truth assignment $\TA$ for $\psi$:
    (i) assign $y_1,...,y_k$ some values such that that 
    $\TA(y_1) \X ... \X \TA(y_k) \neq \parity{}$,
    (ii) assign the other non-basic variables in $\Eqs$ with arbitrary values,
    and
    (iii) evaluate the values of the basic variables.
    Thus it is not possible that
    $\psi \Models (y_1 \X ... \X y_k \Equal \parity{})$
    and
    the case of $\XCB \LinComb \XCB'$ equaling to
    $y_1 \X ... \X y_k \Equal \parity{}$ is impossible.
  \end{itemize}
\end{itemize}
\end{IEEEproof}
\else 
\begin{IEEEproof}
  There are two cases to consider.
  \begin{itemize}
  \item
  Case I: $\psi$ is unsatisfiable.

  If there is a subset $S$ of xor-constraints in $\psi$ such that
  $\bigoplus_{\XC \in S} \XC = (\F \Equal \T)$,
  then,
  by iteratively applying ${\XCI1 \land \XCI2} \Models {\XCI1 \land \XCI2 \land (\XCI1 \X \XCI2)}$,
  we have $\bigwedge_{\XC \in S} \XC \Models \bigoplus_{\XC \in S} \XC$,
  i.e.~$\bigwedge_{\XC \in S} \XC \Models (\F \Equal \T)$,
  and
  thus $\psi$ is unsatisfiable.

  For the other direction, assume that $\psi$ is unsatisfiable.
  Represent the conjunction $\psi$ as a system of linear equation
  modulo two in matrix form. Gaussian elimination must
  result in an equation $ 0 \equiv 1 \mod 2 $
  in some row $r$ of the matrix. The row $r$ is a linear combination
  of the rows $ \set{r_1, \dots, r_n}$. Each row $r_i$ in the matrix
  corresponds to a distinct xor-constraint $C(r_i)$ in $\psi$.
  Thus, $ S = \set{C(r_1), \dots, C(r_n)} \subseteq \psi$
  is a subset of xor-constraints in $\psi$ such that
  $\bigoplus_{\XC \in S} \XC = (\F \Equal \T)$.

  \item
  Case II: $\psi$ is satisfiable.

  If there is a subset $S$ of xor-constraints in $\psi$ such that
  $\bigoplus_{\XC \in S} \XC = \XCB$,
then,   by iteratively applying ${\XCI1 \land \XCI2} \Models {\XCI1 \land \XCI2 \land (\XCI1 \X \XCI2)}$,
we have $\bigwedge_{\XC \in S} \XC \Models \bigoplus_{\XC \in S} \XC$
and   thus $\bigwedge_{\XC \in S} \XC \Models \XCB$ and $\psi \Models \XCB$.

Assume that $\psi \Models \XCB$ for an xor-constraint $\XCB = y_1 \oplus ... \oplus y_n \equiv p$. We have
$\VarsOf{\XCB} \subseteq \VarsOf{\psi}$. Create a (reduced row echelon form)
tableau $\Eqs$ of $\psi$ such that the maximum number of variables of $\XCB$
are chosen to be basic variables. Let $V_b \subseteq \VarsOf{\XCB} $ denote the set of basic variables of $\XCB$ in $\Eqs$. Note that each equation $ \XC $ in
$\Eqs$ is a linear combination of a subset $S(\XC)$ of xor-constraints in
$\psi$. Consider the linear combination $\XCB' = \bigoplus_{\XC \in \Eqs
\wedge \VarsOf{\XC} \cap V_b \not = \emptyset} \XC $ of each equation in $\Eqs$
having at least one common variable with $\XCB$. There are two cases to consider:
\begin{itemize}
\item Case (i): each variable of $\XCB$ is a basic variable in $\Eqs$.  It
clearly holds that $\VarsOf{\XCB} \subseteq \VarsOf{\XCB'}$ and $\psi \models \XCB'$. It
remains to show that the xor-constraint $\XCB'$ does not have
other variables. Assume, for the sake of contradiction, that $\XCB'$
has a variable $x \in \VarsOf{\Eqs} \backslash \VarsOf{\XCB}$, e.g. $ \XCB' = y_1 \oplus ... \oplus y_n \oplus x \equiv p' $. The
variable $x$ cannot be a basic variable because in each selected equation
a variable in $\XCB$ is a basic variable, so $x$ must be a non-basic variable.
A satisfying truth assignment for $\Eqs$ can be constructed by assigning
arbitrary values for non-basic variables and then evaluating the values of
basic variables. Since $\Eqs$ is satisfiable, there must be two satisfying truth
assignments $\TA$ and $\TA'$ such that $\TA(x) = \top$ and $\TA'(x) = \bot$,
            meaning that both xor-constraints $ y_1 \oplus ... \oplus y_n
            \oplus \top \equiv p'$ and $y_1 \oplus ... \oplus y_n \oplus \bot \equiv p'$
            must be satisfiable. Since $ \psi \models y_1 \oplus ... \oplus y_n
            \equiv p $, this is not possible, so $\VarsOf{\XCB'} = \VarsOf{\XCB}$ and $p =
            p'$. 
\item Case (ii): at least one variable of $\XCB$ is a non-basic variable in $\Eqs$.
            Each basic variable of $\XCB$ has exactly one occurrence in $\Eqs$, so $\XCB'$ clearly contains all basic variables of $\XCB$. Assume, for the sake of contradiction, that $\VarsOf{\XCB}\backslash \VarsOf{\XCB'}$ contains a non-empty set $X$ of non-basic variables. Since $\psi \Models \XCB' \wedge \XCB$, it follows that $\psi \Models \bigoplus_{x \in X} x \equiv p''$ where $p'' \in \set{\top,\bot}$. However, since all variables in $X$ are non-basic, they can be assigned to any value. There is thus a truth assignment $\TA$ such that $\TA \Models \psi$ and $\TA \not \Models \bigoplus_{x \in X} x \equiv p''$. This is not possible, so $\VarsOf{\XCB} \subseteq \VarsOf{\XCB'}$. Now, the proof
                 proceeds as in case (i).
\end{itemize} 
It remains to show that a subset $S \subseteq \psi$ such that $\bigwedge_{\XC
\in S} \XC \Models \XCB$ exists.
Each equation $\XC$ in $\Eqs$ is a linear combination of the subset $ S(\XC)
    \subseteq \psi$ and the xor-constraint $E$ is a linear combination of the
    rows in $\Eqs$ that contain basic variables of $E$. In case an xor-constraint in $\psi$ occurs an even number of times in these rows, it cannot be in the set $S$. The subset $S \subseteq \psi$ is formed by taking the xor-constraints with an odd number of occurrences in the family of sets 
     $ \set{ S(\XC) |\; \XC \in \Eqs \wedge  \VarsOf{\XC} \cap V_b \not = \emptyset} $.

  \end{itemize}%
\end{IEEEproof}
\fi 

Another key property of tableaux is that the equations in them are logical
consequences of the represented conjunction of xor-constraints:
\begin{fact}\label{Fact:Tableau}
  If $\Eqs$ is a tableau for $\xorpart$,
  then
  $\Var_i \Def \Var_{i,1} \X ... \X \Var_{i,k_i} \X \parity{i} \in \Eqs$
  implies
  $\xorpart \Models (\Var_i \X \Var_{i,1} \X ... \X \Var_{i,k_i} \Equal \parity{i})$.
\end{fact}

%
%
\subsection{Proof of Lemma~\ref{Lemma:SimplexConsistent}}

\begin{relemma}{\ref{Lemma:SimplexConsistent}}
  Let $\Tuple{\Eqs,\TA}$ be a propagation saturated assigned tableau for
  $\xorpart$.
  %
  %
  The formula $\xorpart \land \bigwedge_{(x \mapsto v) \in \TA}(x \Equal v)$
  is satisfiable
  if and only if
  $\Tuple{\Eqs,\TA}$ is consistent.
\end{relemma}
\begin{IEEEproof}
  First, assume that $\Tuple{\Eqs,\TA}$ is consistent.
  Extend the assignment $\TA$ into a total one $\TA'$ by
  (i) assigning arbitrary values to the unassigned non-basic variables,
  and
  (ii) evaluating the unassigned basic variables according
  to their equations.
  As $\Tuple{\Eqs,\TA}$ is propagation saturated and consistent,
  the resulting truth assignment $\TA'$ does not violate any of the equations.
  Because
  $\bigwedge_{{\Var_i \Def \Var_{i,1} \X ... \X \Var_{i,k_i} \X \parity{i}} \in \Eqs} (\Var_i \X \Var_{i,1} \X ... \X \Var_{i,k_i} \Equal \parity{i})$
  is logically equivalent to $\xorpart$,
  formula $\xorpart \land \bigwedge_{(x \mapsto v) \in \TA}(x \Equal v)$
  is satisfied by $\TA'$.

  Now, assume that $\Tuple{\Eqs,\TA}$ is inconsistent.
  Then there is an equation
  $\Var_i \Def {\Var_{i,1} \X ... \X \Var_{i,k_i} \X \parity{i}}$
  such that
  $\TA(x)$ is defined for all $x \in \Set{\Var_i,\Var_{i,1},...,\Var_{i,k_i}}$
  and
  $\TA(\Var_i) \neq {\TA(\Var_{i,1}) \X ... \X \TA(\Var_{i,k_i}) \X \parity{i}}$.
  By Fact.~\ref{Fact:Tableau},
  $\xorpart \Models (\Var_i \X \Var_{i,1} \X ... \X \Var_{i,k_i} \Equal \parity{i})$.
  Now $\TA$ or any of its extensions do not satisfy
  $(\Var_i \X \Var_{i,1} \X ... \X \Var_{i,k_i} \Equal \parity{i})$;
  thus $\TA$ or any of its extensions do not satisfy $\xorpart$ either.
  As a result,
  the formula $\xorpart \land \bigwedge_{(x \mapsto v) \in \TA}(x \Equal v)$
  is unsatisfiable.
\end{IEEEproof}

%
%
\subsection{Proof of Lemma~\ref{Lemma:SimplexImpliedUna}}

\begin{relemma}{\ref{Lemma:SimplexImpliedUna}}
  Let $\Tuple{\Eqs,\TA}$ be a consistent,
  propagation saturated assigned tableau for $\xorpart$.
  For each literal $y \Equal v_y$ it holds that
  ${\xorpart \land  \bigwedge_{(x \mapsto v_x) \in \TA}(x \Equal v_x)} \Models (y \Equal v_y)$
  if and only if
  $\TA(y) = v_y$.
\end{relemma}
\begin{IEEEproof}
  If $\TA(y) = v_y$,
  then ${\xorpart \land \bigwedge_{(x \mapsto v_x) \in \TA}(x \Equal v_x)} \Models (y \Equal v_y)$ holds trivially as
  $(y \mapsto v_y) \in \TA$.

  Assume that ${\xorpart \land  \bigwedge_{(x \mapsto v_x) \in \TA}(x \Equal v_x)} \Models (y \Equal v_y)$ holds.
  As $\Tuple{\Eqs,\TA}$ is consistent and propagation saturated,
  by Lemma~\ref{Lemma:SimplexConsistent}
  $\xorpart \land  \bigwedge_{(x \mapsto v_x) \in \TA}(x \Equal v_x)$
  is satisfiable.
  Suppose that $\Eqs$ has $n$ non-basic variables not assigned by $\TA$.
  As $\Tuple{\Eqs,\TA}$ is consistent and propagation saturated,
  there are $2^n$ total extensions of $\TA$ that respect the equations in $\Eqs$,
  obtained by assigning arbitrary values to the unassigned non-basic variables
  and then evaluating the unassigned basic variables.
  All these extensions satisfy
  $\xorpart \land  \bigwedge_{(x \mapsto v_x) \in \TA}(x \Equal v_x)$.
  For each $\TA$-unassigned variable $y$ there is thus at least one
  satisfying truth assignment where $y$ is $\F$ and one where $y$ is $\T$.
  Thus ${\xorpart \land  \bigwedge_{(x \mapsto v_x) \in \TA}(x \Equal v_x)} \Models (y \Equal v_y)$ can hold only if
  $y$ is assigned by $\TA$ and $\TA(y) = v_y$.
\end{IEEEproof}

\subsection{Proof of Lemma~\ref{Lemma:SimplexImpliedBin}}

\newcommand{\TAExts}{\Gamma}

\begin{relemma}{\ref{Lemma:SimplexImpliedBin}}
  Let $\Tuple{\Eqs,\TA}$ be a consistent,
  propagation saturated assigned tableau for $\xorpart$.
  For any two distinct variables $y,z$ and any $\parity{} \in \Booleans$,
  it holds that
  ${\xorpart \land  \bigwedge_{(x \mapsto v_x) \in \TA}(x \Equal v_x)} \Models (y \X z \Equal \parity{})$
  if and only if
  \begin{enumerate}
  \item
    $\TA(y)$ and $\TA(z)$ are both defined and
    ${\TA(y) \X \TA(z) = \parity{}}$,
  \item
    $\TA(y)$ and $\TA(z)$ are undefined and
    $\Eqs$ has an equation $\Eq$ of form $y \Def {... \X z \X ...}$
    such that $\Restr{\Eq}{\TA}$ is ${y \Def z \X \parity{}}$,
    where $\Restr{\Eq}{\TA}$ is the equation obtained from $\Eq$ by substituting
    the variables in it assigned by $\TA$ with their values,
  \item
    $\TA(y)$ and $\TA(z)$ are undefined and
    $\Eqs$ has an equation $\Eq$ of form $z \Def {... \X y \X ...}$
    such that $\Restr{\Eq}{\TA}$ is ${z \Def y \X \parity{}}$,
    or
  \item
    $\TA(y)$ and $\TA(z)$ are undefined and
    $\Eqs$ has two equations, $\Eq_y$ and $\Eq_z$,
    of forms $y \Def ...$ and $z \Def ...$
    such that
    $\Restr{\Eq_y}{\TA}$ is $y \Def f$,
    $\Restr{\Eq_z}{\TA}$ is $z \Def g$,
    and
    $f \X g$ equals $\parity{}$.
  \end{enumerate}
\end{relemma}
\begin{IEEEproof}
  Because $\Tuple{\Eqs,\TA}$ is consistent and propagation saturated,
  there are $2^n$ total extensions of $\TA$ that respect all the equations in $\Eqs$,
  obtained by assigning arbitrary values to the $n$ $\TA$-unassigned non-basic variables in $\Tuple{\Eqs,\TA}$ and
  then evaluating the $\TA$-unassigned basic variables according to the equations.
  In the following,
  the set of all such extensions is denoted by $\TAExts$.
  Furthermore,
  all such total extensions also satisfy
  $\xorpart \land \bigwedge_{(x \mapsto v_x) \in \TA}(x \Equal v_x)$
  because
  $\bigwedge_{{\Var_i \Def \Var_{i,1} \X ... \X \Var_{i,k_i} \X \parity{i}} \in 
    \Eqs} (\Var_i \X \Var_{i,1} \X ... \X \Var_{i,k_i} \Equal \parity{i})$
  is logically equivalent to $\xorpart$.
  For the same reason,
  they are also the only truth assignments over $\VarsOf{\xorpart}$ that satisfy
  $\xorpart \land \bigwedge_{(x \mapsto v_x) \in \TA}(x \Equal v_x)$.
  And for each $\TA$-unassigned variable $x$,
  there is an extension $\TA' \in \TAExts$ with $\TA'(x) = \F$ and
  another extension $\TA'' \in \TAExts$ with $\TA''(x) = \T$.

  First,
  assume that 
  $\TA(y)$ and $\TA(z)$ are both defined.
  Now it is straightforward to observe that
  ${\TA(y) \X \TA(z) = \parity{}}$ if and only if
  ${\xorpart \land  \bigwedge_{(x \mapsto v_x) \in \TA}(x \Equal v_x)} \Models (y \X z \Equal \parity{})$.

  Second,
  assume that $\TA(y)$ is defined but $\TA(z)$ is not
  (the case when $\TA(z)$ is defined but $\TA(y)$ is not is symmetric to this).
  %
  %
  %
  %
  As $z$ is $\TA$-unassigned,
  there is a $\TA' \in \TAExts$ with $\TA'(z)=\F$
  and
  a $\TA'' \in \TAExts$ with $\TA'(z)=\T$.
  As $\TA'$ and $\TA''$ also satisfy
  $\xorpart \land  \bigwedge_{(x \mapsto v_x) \in \TA}(x \Equal v_x)$,
  ${\xorpart \land  \bigwedge_{(x \mapsto v_x) \in \TA}(x \Equal v_x)} \Models (y \X z \Equal \parity{})$ cannot hold.

  Lastly, assume that $\TA(y)$ and $\TA(z)$ are both undefined.
  We have four cases to consider.
  \begin{enumerate}
  \item
    $y$ and $z$ are both non-basic variables.
    Now there are extensions $\TA_1,\TA_2,\TA_3,\TA_4 \in \TAExts$
    covering all the four truth value combinations possible for
    the variable pair $y$ and $z$.
    As all these also satisfy
    $\xorpart \land  \bigwedge_{(x \mapsto v_x) \in \TA}(x \Equal v_x)$,
    ${\xorpart \land  \bigwedge_{(x \mapsto v_x) \in \TA}(x \Equal v_x)} \Models (y \X z \Equal \parity{})$ cannot hold.

  \item
    $y$ is a basic variable and $z$ is a non-basic variable.

    Take the equation $\Eq$ of form $y \Def ...$ for $y$ in $\Eqs$.
    As $\Tuple{\Eqs,\TA}$ is consistent and propagation saturated,
    and $y$ is $\TA$-unassigned,
    there is at least one $\TA$-unassigned variable in the right hand side of $\Eq$.

    If $\Restr{\Eq}{\TA}$ is $y \Def z \X \parity{}$ for some $\parity{} \in \Booleans$,
    i.e.~there is exactly one $\TA$-unassigned variables in the right hand side of $\Eq$ and that variable is $z$,
    then
    the value of $y$ is fully determined by the value of $z$ in each $\TA' \in \TAExts$,
    and thus
    ${\xorpart \land  \bigwedge_{(x \mapsto v_x) \in \TA}(x \Equal v_x)} \Models (y \X z \Equal \parity{})$ holds.
    
    On the other hand,
    if ${\xorpart \land  \bigwedge_{(x \mapsto v_x) \in \TA}(x \Equal v_x)} \Models (y \X z \Equal \parity{})$ holds,
    then the value of $y$ is fully determined by the value of $z$ in each $\TA' \in \TAExts$
    and
    thus $\Restr{\Eq}{\TA}$ must be $y \Def z \X \parity{}$.

  \item
    $z$ is a basic variable and $y$ is a non-basic variable.
    This case is symmetric to the previous one.

  \item
    $y$ and $z$ are both basic variables.

    If $\Eqs$ has two equations, $\Eq_y$ and $\Eq_z$,
    of forms $y \Def ...$ and $z \Def ...$
    such that
    $\Restr{\Eq_y}{\TA}$ is $y \Def f$,
    $\Restr{\Eq_z}{\TA}$ is $z \Def g$,
    and
    $f \X g$ (with duplicate variables eliminated) equals $\parity{}$,
    then $f$ and $g$ must contain the same variables as otherwise $f \X g$ would not equal $\parity{}$.
    Thus $\Eq_y$ and $\Eq_z$ must contain the same $\TA$-unassigned variables
    and
    consequently $\TA'(y) = \TA'(z) \X \parity{}$ for each $\TA' \in \TAExts$,
    implying
    ${\xorpart \land  \bigwedge_{(x \mapsto v_x) \in \TA}(x \Equal v_x)} \Models (y \X z \Equal \parity{})$.
    
    If ${\xorpart \land  \bigwedge_{(x \mapsto v_x) \in \TA}(x \Equal v_x)} \Models (y \X z \Equal \parity{})$ holds,
    then
    the equations $\Eq_y$ and $\Eq_z$ for $y$ and $z$, resp.,
    must contain the same $\TA$-unassigned non-basic variables
    because otherwise there would be extensions $\TA',\TA'' \in \TAExts$
    such that $\TA'(y)=\TA''(y)$ but $\TA'(z)\neq\TA''(z)$ and
    ${\xorpart \land  \bigwedge_{(x \mapsto v_x) \in \TA}(x \Equal v_x)} \Models (y \X z \Equal \parity{})$ would not hold.
    As a consequence,
    $\Restr{\Eq_y}{\TA}$ is $y \Def f$,
    $\Restr{\Eq_z}{\TA}$ is $z \Def g$,
    and
    $f \X g$ equals $\parity{}$.
  \end{enumerate}%
\end{IEEEproof}

%
%
\subsection{Proof of the Decomposition Theorem~\ref{Thm:Decomposition}}

\newcommand{\VAp}{V'_\textup{a}}
\newcommand{\VBp}{V'_\textup{b}}

\begin{retheorem}{\ref{Thm:Decomposition}}
  Let $(\VA,\VB)$ be an $\Var$-cut partition of $\xorpart$.
  Let $\xorpartA = \bigwedge_{\XC \in \VA}\XC$,
  $\xorpartB = \bigwedge_{\XC \in \VB}\XC$, and
  $\AL_1,...,\AL_k,\IL \in\LitsOf{\xorpart}$.
  Then it holds that:
  \begin{itemize}
  \item
    If $\xorpart \land {\AL_1 \land ... \land \AL_k}$ is unsatisfiable,
    then
    \begin{enumerate}
    \item
      $\xorpartA \land {\AL_1 \land ... \land \AL_k}$ or
      $\xorpartB \land {\AL_1 \land ... \land \AL_k}$ is unsatisfiable;
      or
    \item
      $\xorpartA \land {\AL_1 \land ... \land \AL_k} \Models (\Var \Equal \parity{\Var})$ and $\xorpartB \land {\AL_1 \land ... \land \AL_k} \Models (\Var \Equal \parity{\Var} \X \T)$ for some $\parity{\Var} \in \Set{\F,\T}$.
    \end{enumerate}
  \item
    If $\xorpart \land {\AL_1 \land ... \land \AL_k}$ is satisfiable
    and
    $\xorpart \land {\AL_1 \land ... \land \AL_k} \Models \IL$,
    then
    \begin{enumerate}
    \item
      $\xorpartA \land {\AL_1 \land ... \land \AL_k} \Models \IL$ or
      $\xorpartB \land {\AL_1 \land ... \land \AL_k} \Models \IL$; or
    \item
      $\xorpartA \land {\AL_1 \land ... \land \AL_k} \Models (\Var \Equal \parity{\Var})$ and $\xorpartB \land {\AL_1 \land ... \land \AL_k \land (\Var \Equal \parity{\Var})} \Models \IL$;
      or
    \item
      $\xorpartB \land {\AL_1 \land ... \land \AL_k} \Models (\Var \Equal \parity{\Var})$ and $\xorpartA \land {\AL_1 \land ... \land \AL_k \land (\Var \Equal \parity{\Var})} \Models \IL$.
    \end{enumerate}
  \end{itemize}
\end{retheorem}
\begin{IEEEproof}
  Let $(\VAp,\VBp)$ be an $\Var$-cut partition of
  $\xorpart \land (\AL_1) \land ... \land (\AL_k)$
  with
  $\VarsOf{\VAp} = \VarsOf{\VA}$,
  $\VarsOf{\VBp} = \VarsOf{\VB}$,
  $\VA \subseteq \VAp$, and
  $\VB \subseteq \VBp$.
  Such partition exists because the xor-assumption literals $\AL_i$
  are unit xor-constraints.

  Case I: $\xorpart \land {\AL_1 \land ... \land \AL_k}$ is unsatisfiable.
  By Lemma~\ref{Lemma:LinearCombs},
  there is a subset $S$ of xor-constraints
  in $\xorpart \land (\AL_1) \land ... \land (\AL_k)$
  such that $\BigLinComb_{\XC \in S} \XC = (\F \Equal \T)$.
  Observe that
  $\BigLinComb_{\XC \in S} \XC =
   (\BigLinComb_{\XC \in {\VAp \cap S}} \XC) \LinComb
   (\BigLinComb_{\XC \in {\VBp \cap S}} \XC)$.
  If $\BigLinComb_{\XC \in {\VAp \cap S}} \XC = (\F \Equal \T)$,
  then
  $\xorpartA \land {\AL_1 \land ... \land \AL_k}$ is also unsatisfiable.
  Similarly,
  if $\BigLinComb_{\XC \in {\VBp \cap S}} \XC = (\F \Equal \T)$,
  then $\xorpartB \land {\AL_1 \land ... \land \AL_k}$ is unsatisfiable.
  Otherwise,
  it must be that
  $\BigLinComb_{\XC \in {\VAp \cap S}} \XC = (\Var \Equal \parity{\Var})$
  and
  $\BigLinComb_{\XC \in {\VBp \cap S}} \XC = (\Var \Equal \parity{\Var} \X \T)$
  with $\parity{\Var} \in \Set{\F,\T}$
  because
  $\VAp \cap \VBp = \emptyset$,
  ${\VarsOf{\VAp} \cap \VarsOf{\VBp}} = \Set{\Var}$ and
  $(\BigLinComb_{\XC \in {\VAp \cap S}} \XC) \LinComb
   (\BigLinComb_{\XC \in {\VBp \cap S}} \XC) = (\F \Equal \T)$.
  Thus
  $\xorpartA \land {\AL_1 \land ... \land \AL_k} \Models (\Var \Equal \parity{\Var})$ and
  $\xorpartB \land {\AL_1 \land ... \land \AL_k} \Models (\Var \Equal \parity{\Var} \X \T)$.
  
  Case II:
  $\xorpart \land \AL_1 \land ... \land \AL_k$ is satisfiable
  and
  $\xorpart \land \AL_1 \land ... \land \AL_k \Models \IL$
  with
  $\IL = (y \Equal \parity{y})$ for some variable $y$ and
  $\parity{y} \in \Set{\F,\T}$.
  There is a subset $S$ of xor-constraints
  in $\xorpart \land (\AL_1) \land ... \land (\AL_k)$
  such that ${\BigLinComb_{\XC \in S} \XC} = (y \Equal \parity{y})$.
  Again,
  observe that
  $(\BigLinComb_{\XC \in S} \XC) = 
   (\BigLinComb_{\XC \in {\VAp \cap S}} \XC) \LinComb
   (\BigLinComb_{\XC \in {\VBp \cap S}} \XC)$
  and
  thus it must be that either
  $y \in \VarsOf{\BigLinComb_{\XC \in {\VAp \cap S}} \XC}$ or
  $y \in \VarsOf{\BigLinComb_{\XC \in {\VBp \cap S}} \XC}$
  but not both.
  Assume that $y \in \VarsOf{\BigLinComb_{\XC \in {\VBp \cap S}} \XC}$;
  the other case is symmetric.
  Now
  $\VarsOf{\BigLinComb_{\XC \in {\VAp \cap S}} \XC} \subseteq \Set{\Var}$
  and
  $\VarsOf{\BigLinComb_{\XC \in {\VBp \cap S}} \XC} \subseteq \Set{\Var,y}$.
  If $\Var \in \VarsOf{\BigLinComb_{\XC \in {\VAp \cap S}} \XC}$,
  then
  $\BigLinComb_{\XC \in {\VAp \cap S}} \XC = (\Var \Equal \parity{\Var})$
  for a $\parity{\Var} \in \Set{\F,\T}$,
  $\xorpartA \land {\AL_1 \land ... \land \AL_k} \Models (\Var \Equal \parity{\Var})$,
  $\BigLinComb_{\XC \in {\VBp \cap S}} \XC = (\Var \X y \Equal \parity{\Var} \X \parity{y})$,
  and
  $\xorpartB \land {\AL_1 \land ... \land \AL_k \land (\Var \Equal \parity{\Var})} \Models {y \X \parity{y}}$.
  If $\Var \notin \VarsOf{\BigLinComb_{\XC \in {\VAp \cap S}} \XC}$,
  then
  $\Var \notin \VarsOf{\BigLinComb_{\XC \in {\VBp \cap S}} \XC}$,
  $\BigLinComb_{\XC \in {\VBp \cap S}} \XC = (y \Equal \parity{y})$,
  and
  $\xorpartB \land {\AL_1 \land ... \land \AL_k} \Models (y \Equal \parity{y})$.
\end{IEEEproof}



\end{document}